\documentclass[reprint, showpacs, superscriptaddress, aip, pof, longbibliography, twocolumn, 10pt]{revtex4-1}
\usepackage{graphicx}
\usepackage{color}
\usepackage{amssymb}
\usepackage{amsmath}
\usepackage{tabularx}
\usepackage{bm}
\usepackage{hyperref}
\usepackage{ltxtable}
\usepackage{natbib} 
\usepackage{longtable}
\usepackage{float}

\newcommand{\be}{\begin{equation}}
\newcommand{\ee}{\end{equation}}
\newcommand{\bse}{\begin{subequations}}
	\newcommand{\ese}{\end{subequations}}
\newcommand{\bea}{\begin{eqnarray}}
\newcommand{\eea}{\end{eqnarray}}
\newcommand{\ba}{\begin{array}}
	\newcommand{\ea}{\end{array}}

\usepackage[usenames,dvipsnames]{xcolor}
\hypersetup{
colorlinks,
citecolor=blue,
filecolor=blue,
linkcolor=blue,
urlcolor=blue}

\begin{document}

\title{Predictions of Reynolds and Nusselt numbers in turbulent convection using machine-learning models}
\author{Shashwat Bhattacharya}
\email{shashwat.bhattacharya@tu-ilmenau.de}
\affiliation{Institut f{\"ur} Thermo-und Fluiddynamik, Technische Universit{\"a}t Ilmenau, Postfach 100565, D-98684 Ilmenau, Germany}
\author{Mahendra K. Verma}
\affiliation{Department of Physics, Indian Institute of Technology Kanpur, Kanpur 208016, India}
\author{{Arnab Bhattacharya}}
\affiliation{Department of Computer Science and Engineering, Indian Institute of Technology Kanpur, Kanpur 208016, India}

\date{\today}

\begin{abstract}
In this paper, we develop a multivariate regression model and a neural network model to predict the Reynolds number (Re) and Nusselt number in turbulent thermal convection. We compare  their predictions with those of earlier models of convection:  Grossmann-Lohse~[Phys. Rev. Lett. \textbf{86}, 3316 (2001)], revised Grossmann-Lohse~[Phys. Fluids \textbf{33}, 015113 (2021)], and Pandey-Verma [Phys. Rev. E \textbf{94}, 053106 (2016)] models. We observe that although the predictions of all the models are quite close to each other, {\color{black}the machine learning models developed in this work provide the best match with the experimental and numerical results. }
\end{abstract}
\maketitle

\section{Introduction}
\label{sec:Introduction}
Thermal convection is encountered in natural phenomena, such as stars and planets, as well as in various engineering applications. A simplified version of thermal convection is Rayleigh-B{\'e}nard convection (RBC), which is a model of a fluid enclosed between two horizontal plates with the bottom plate kept at a higher temperature than the top one.  RBC is governed primarily by two nondimensional parameters: the Rayleigh number Ra, which is the ratio of the buoyancy  and the dissipative force, and the Prandtl number Pr, which is the ratio of kinematic viscosity and thermal diffusivity of the fluid~\cite{Chandrasekhar:book:Instability,Ahlers:RMP2009,Chilla:EPJE2012}. Two important response parameters of RBC are the Nusselt number (Nu) and the Reynolds number (Re), which are respective measures of large-scale heat transport and velocity in turbulent RBC~\cite{Ahlers:RMP2009,Chilla:EPJE2012}. The Reynolds number is defined as $\mathrm{Re}= Ud/\nu$,
where $U$ is the large-scale velocity, $\nu$ is the kinematic viscosity, and $d$ is the distance between the thermal plates. The Nusselt number is given by $\mathrm{Nu} = 1 + \langle u_zT \rangle/(\kappa \Delta d^{-1})$, where $u_z$ is the vertical component of velocity, $T$ is the temperature field, $\kappa$ is the thermal diffusivity,  $\Delta$ is the temperature difference between the thermal plates, and $\langle~\rangle$ denotes the volume average.

Researchers have attempted to  model Re and Nu on the governing parameters, Ra and Pr~\cite{Ahlers:RMP2009,Chilla:EPJE2012,Siggia:ARFM1994,Xia:TAML2013,Verma:book:BDF}.
Early theoretical, numerical, and experimental studies of RBC reveal a power-law scaling of Nu and Re, i.e.,  $\mathrm{Nu} \sim \mathrm{Ra}^{\alpha}\mathrm{Pr}^{\beta}$ and $\mathrm{Re} \sim \mathrm{Ra}^{\gamma}\mathrm{Pr}^{\delta}$. The exponents $\alpha, \beta, \gamma$, and $\delta$   vary for different regimes of Pr and Ra. For the scaling of Nu, the exponent $\alpha$ ranges from 1/4 for $\mathrm{Pr} \ll 1$ to approximately 1/3 for $\mathrm{Pr} \gtrsim 1$,~\cite{Malkus:PRSA1954,Shraiman:PRA1990,Cioni:JFM1997,Scheel:PRF2017,Castaing:JFM1989,Chavanne:PRL1997,Horn:JFM2013,Wagner:PF2013,Kaczorowski:JFM2013,Niemela:JFM2003,Funfschilling:JFM2005,Stevens:JFM2010,Dong:PF2020,Madanan:PF2020,Vial:PF2017}. However, Nu has a  weak dependence on Pr, with the exponent $\beta$ ranging from approximately zero for $\mathrm{Pr}\gtrsim 1$ to $0.14$ for small $\mathrm{Pr}$~\cite{Xia:PRL2002,Verzicco:JFM1999}.  Regarding the scaling of Re, the exponent $\gamma$ was observed to be approximately $2/5$ for $\mathrm{Pr}\ll 1$, $1/2$ for $\mathrm{Pr} \sim 1$, and $3/5$ for $\mathrm{Pr} \gg 1$~\cite{Castaing:JFM1989,Chavanne:PRL1997,Cioni:JFM1997,Niemela:JFM2001,Lam:PRE2002,Emran:JFM2008,Silano:JFM2010,Verma:PRE2012,Wagner:PF2013,Scheel:PRF2017,Horn:JFM2013,Pandey:PF2016,Pandey:PRE2016}; and $\delta$ has been observed to range from $ -0.7 $ for $\mathrm{Pr} \lesssim 1$ to $ -0.95 $ for $\mathrm{Pr} \gg 1$.~\cite{Xia:PRL2002,Brown:JSM2007}.
In the limit of very large Ra (the ultimate regime), Kraichnan~\cite{Kraichnan:PF1962Convection} argued that $\mathrm{Nu} \sim \sqrt{\mathrm{RaPr}}$, $\mathrm{Re} \sim \sqrt{\mathrm{Ra/Pr}}$ for $\mathrm{Pr} \leq 0.15$, and $\mathrm{Nu} \sim \sqrt{\mathrm{RaPr^{-1/2}}}$, $\mathrm{Re} \sim \sqrt{\mathrm{Ra/Pr^{3/2}}}$ for $0.15 < \mathrm{Pr} \leq 1$, with logarithmic corrections.
However, the  existence of the aforementioned regime in RBC is still under debate~\cite{Verma:PRE2012,Lohse:PRL2003,Pawar:PRF2016,Chavanne:PRL1997,Schmidt:JFM2012,Roche:PRE2001,Ahlers:NJP2012,He:PRL2012,Niemela:Nature2000,Urban:PRL2012}. 

{\color{black}Since the scaling of Re and Nu depends crucially on the regime of Ra and Pr, researchers felt a need for a unified model that encompasses all the regimes.} Grossmann and Lohse~\cite{Grossmann:JFM2000,Grossmann:PRL2001} (henceforth referred to as GL)  derived the first set of formulae for Re and Nu.  Recently, \citet{Bhattacharya:PF2021} revised the GL model by incorporating  the observed suppression in the dissipation rates in turbulent convection~\cite{Emran:JFM2008,Pandey:PF2016,Pandey:PRE2016,Bhattacharya:PF2018,Bhattacharya:PF2019}. Pandey and Verma~\cite{Pandey:PF2016,Pandey:PRE2016} also derived formulae for Nu and Re. All the above models have been largely successful in predicting Nu and Re over a range of Ra and Pr.

In this paper, we employ \textit{machine-learning} (ML) tools to construct models for predicting Re and Nu. This approach, which is increasingly gaining popularity in fluid mechanics~\cite{Parish:JCP2016,Fonda:PNAS2019,Pandey:PRF2020,Pandey:JOT2020,Brunton:ARFM2020}, involves building and improving prediction algorithms by ``learning" from the existing data~\cite{Goodfellow:book,Hastie:book,Burkov:book}. Hence, in the present work, we do not delve much into the physics behind Re and Nu relations; instead, we use the data from previous works to develop the prediction models.  We construct a \textit{multivariate regression} (MR) model and a \textit{neural network} (NN) model for the predictions of Nu and Re.   
We compare the predictions of the above ML models with those of the earlier models.  We observe that the predictions of all the models are  close to each other, but {\color{black}the machine learning models developed in this study provide better} match with the experimental and numerical results.

The outline of the paper is as follows. In Sec.~\ref{sec:Existing_models}, we briefly describe GL and Pandey-Verma models. In Sec.~\ref{sec:ML}, we discuss the MR and NN models developed in the present study. We compare the predictions of the above models with the existing ones in Sec.~\ref{sec:Results}. We conclude in Sec.~\ref{sec:Conclusion}.

\section{A summary of existing models} 
\label{sec:Existing_models}
In this section, we summarize GL and Pandey-Verma models briefly.

\subsection{Grossmann-Lohse (GL) model}
\label{subsec:GL}
Grossmann and Lohse~\cite{Grossmann:JFM2000,Grossmann:PRL2001} provided the first unifying model for thermal convection. Grossmann and Lohse~\cite{Grossmann:JFM2000,Grossmann:PRL2001} derived this model by  splitting the total viscous and thermal dissipation rates ($\tilde{D}_{u}$ and $\tilde{D}_T$ respectively) into bulk and the boundary layer (BL) contributions as
\begin{eqnarray}
\tilde{D}_u &=& \tilde{D}_{u, \mathrm{bulk}} + \tilde{D}_{u, \mathrm{BL}},
\label{eq:TotalViscousDissipation} \\
\tilde{D}_T &=& \tilde{D}_{T, \mathrm{bulk}} + \tilde{D}_{T, \mathrm{BL}}.
\label{eq:TotalThermalDissipation}
\end{eqnarray}
Using the Prandtl-Blasius theory of boundary layers, and assuming homogeneous and isotropic turbulence-like properties of dissipation rates in the bulk, Grossmann and Lohse~\cite{Grossmann:JFM2000,Grossmann:PRL2001} expressed the bulk and boundary layer contributions in terms of Re, Nu, and Pr. These contributions were then substituted in the exact relations relating the total dissipation rates with the Nusselt number~\cite{Shraiman:PRA1990}: 
\begin{eqnarray}
\tilde{D}_u &=& \frac{\nu^3}{d^4}\mathrm{(Nu}-1)\frac{\mathrm{Ra}}{\mathrm{Pr}^2},
\label{eq:SS_Viscous} \\
\tilde{D}_T &=& \frac{\kappa \Delta^2}{d^2}\mathrm{Nu}.
\label{eq:SS_Thermal} 
\end{eqnarray}
Following the above procedure, Grossmann and Lohse arrived at general expressions for Nu and Re valid for all Pr and Ra. The formulae for Re and Nu were later updated by \citet{Stevens:JFM2013} using more up-to-date data of RBC. The GL model provides good estimates for the Nusselt and Reynolds numbers for a wide range of Ra and Pr.

\begin{table*}
	\caption{Summary of different models of RBC for predicting the Renolds number (Re) and the Nusselt number (Nu).  Note: SS = Shraiman and Siggia~\cite{Shraiman:PRA1990}, $\epsilon_u$ = viscous dissipation rate,  $\epsilon_T$ = thermal dissipation rate, ML = Machine learning.}
	\label{table:Models}
	\begin{center}
		\begin{tabular}{ccccc} \hline \hline
			Model & Year & Type & Prediction algorithm & Comments \\ 
			
			\hline 
			Grossmann and Lohse  & 2001, & Analytical & SS formulae & $\epsilon_u \sim U^3/d$ \\ 
			(GL)~\cite{Grossmann:JFM2000,Grossmann:PRL2001,Stevens:JFM2013} & updated in 2013& + data &  & $\epsilon_T \sim U\Delta^2/d$\\ \\
			Revised Grossmann- & 2021 & Analytical  & SS formula with  &  $\epsilon_u \sim (U^3/d)f(\mathrm{Ra,Pr})$\\
			Lohse (RGL)~\cite{Bhattacharya:PF2021} & &+& modified expression  & $\epsilon_T \sim (U\Delta^2/d)f(\mathrm{Ra,Pr})$\\
			& & data &for dissipation rates & \\ \\
			Pandey and Verma & 2016 & Analytical & Model momentum & Prefactors for the\\
			(PV)~\cite{Pandey:PF2016,Pandey:PRE2016} & &+& equation  &momentum equation\\ 
			& &data & &terms\\ \\
			Multivariate regression  & Present study & ML + data & Multivariate linear &  \\
			(MR)& & &regression & \\ \\
			Neural network (NN) & Present study & ML + data & Neural networks &  \\ 
			\hline  \hline
			
		\end{tabular}
	\end{center}
\end{table*}
\subsection{Revised Grossmann-Lohse (RGL) model}
\label{subsec:RGL}
Although the GL model has been quite successful in explaining past experimental and numerical results, it based on certain assumptions that are not very accurate for RBC. First, the model assumes that  the viscous and thermal dissipation rates in the bulk scale as $U^3/d$ and $U\Delta^2/d$ (for $\mathrm{Pr} \leq 1$) respectively, as in passive scalar turbulence with open boundaries~\cite{Lesieur:book:Turbulence,Verma:book:ET}. Recent studies, however,  show that these rates get an additional correction of approximately $\mathrm{Ra}^{-0.2}$ for $\mathrm{Pr} \sim 1$~\cite{Emran:JFM2008,Pandey:PF2016,Pandey:PRE2016,Scheel:PRF2017,Bhattacharya:PF2018,Bhattacharya:PF2019}.
The above correction is because of the inhibition of nonlinear interactions due to the presence of walls and buoyancy~\cite{Verma:book:BDF,Pandey:PF2016,Pandey:PRE2016,Bhattacharya:PF2019b}. In addition, the above studies  reveal that the thickness of the viscous boundary layer  in RBC  deviates from $\mathrm{Re}^{-1/2}$, contrary to what is assumed in GL model~\cite{Scheel:JFM2012,Shi:JFM2012,Bhattacharya:PF2018}.

Recently, \citet{Bhattacharya:PF2021} incorporated the above corrections in the dissipation rates and the boundary layer thickness in the equations of the GL model. In the revised model (henceforth referred to as Revised Grossmann-Lohse or RGL model), the Reynolds and Nusselt numbers are obtained by solving a cubic polynomial equation consisting of four functions [$f_i(\mathrm{Ra,Pr})$] that are prefactors for the dissipation rates in the bulk and boundary layers.  The functional forms of the prefactors in the RGL model were determined using multivariate linear regression on 60 sets of simulation data of RBC. \citet{Bhattacharya:PF2021} observed that the predictions of the RGL model were  marginally better than the GL model, especially for extreme Prandtl numbers~\cite{Bhattacharya:PF2021}.

\subsection{Pandey-Verma (PV) model}
\label{subsec:PV}
Pandey and Verma~\cite{Pandey:PF2016,Pandey:PRE2016} used a different approach from the aforementioned models.   To compute the P{\'e}clet number ($\mathrm{Pe} = \mathrm{RePr}$) as a function of Ra and Pr, Pandey and Verma~\cite{Pandey:PF2016,Pandey:PRE2016} constructed a model (henceforth referred to as PV model) by estimating the relative strengths of the various terms of the momentum equation of RBC using data from their numerical simulations.  Based on these estimates, they arrived at a quadratic formula for the P{\'e}clet number. The predictions of Re by the PV model were as accurate as the those by the GL model~\cite{Verma:book:BDF}.  Pandey and Verma~\cite{Pandey:PF2016,Pandey:PRE2016} also derived a general formula for the Nusselt number using correlations.
Note that the relation for Nu was obtained using simulation data for $\mathrm{Pr}\geq 1$ and thus is not accurate for $\mathrm{Pr} \ll 1$ regime as shown later in this paper. 

The main aspects of GL, RGL, and PV models are summarized in Table~\ref{table:Models}. In the next section, we describe the machine-learning models developed in the present study.
\section{Machine-Learning models}
\label{sec:ML}
Having discussed three important models of the past, we now describe two  machine-learning models to determine the Nusselt and Reynolds numbers for given Rayleigh and Prandtl numbers.  The first one, referred to as Multivariate Regression model (MR),  employs linear regression to the available data directly. The second approach, referred to as Neural Network (NN) model, employs artificial neural networks for the predictions. 

To construct the  ML models, we use 62 sets of simulation data as training datasets. Sixty of these datasets were generated by \citet{Bhattacharya:PF2021} using high-resolution direct numerical simulations of RBC. The Rayleigh number for these datasets ranges from $5 \times 10^5$ to $5 \times 10^9$, and the Prandtl number ranges from $0.02$ to $100$. All the above simulations were conducted using the finite-difference code SARAS~\cite{Samuel:JOSS2020,Verma:SNC2020} on a cubical domain of unit dimension with no-slip walls on all the sides.  Isothermal boundary conditions were imposed on the horizontal walls. The grid resolution varied between $257^3$ to $1025^3$, depending on the resolution requirements. For further details on the simulations and their validations, refer to \citet{Bhattacharya:PF2021,Bhattacharya:PRF2021}.  Additionally, we conduct two simulations of RBC for $\mathrm{Pr}=0.02$ with $\mathrm{Ra}=10^5$ and $2 \times 10^5$; these simulations were performed on a $513^3$ grid. {\color{black} Here, we remark that the output parameters — Re and Nu — are global quantities that are based on velocity and heat fluxes that are averaged over the entire domain. Further, Nu and Re used for training the model are not computed for a specific timeframe; rather they are computed for multiple timeframes (ranging from 14 to 263; see \citet{Bhattacharya:PF2021} for details) and then averaged over these timeframes. Thus, we take care of the spatial and temporal fluctuations of quantities that are typically encountered in turbulent flows.} The  parameters of all the data samples, along with their corresponding Re and Nu, are tabulated in Table~\ref{table:Training}. 
\begin{figure}[t]
	\centering
	\includegraphics[scale=0.4]{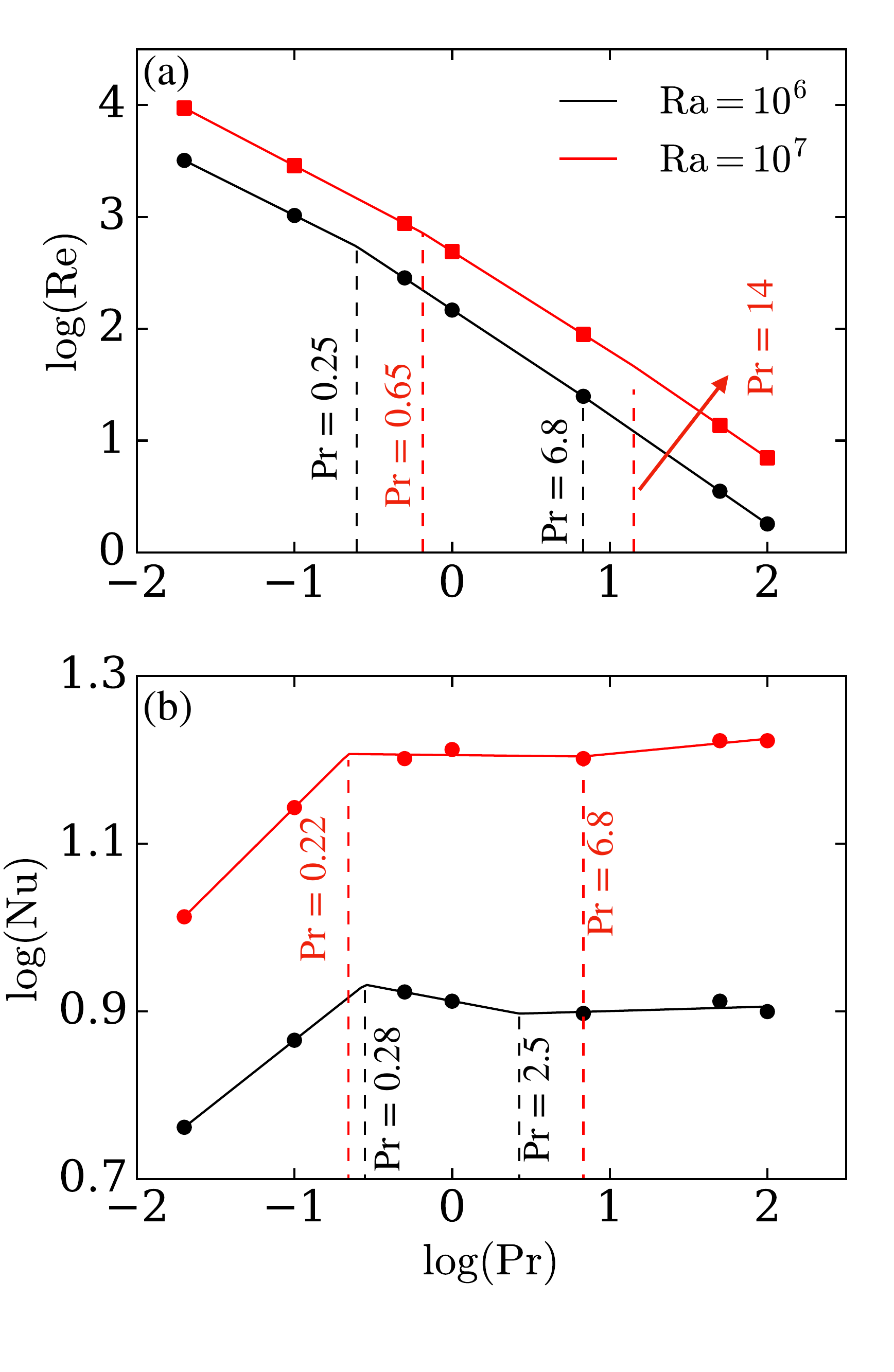}
	\caption{{\color{black}(color online) For $\mathrm{Ra}=10^6$ (black) and $\mathrm{Ra}=10^7$ (red): (a) plots of Re versus Pr, and (b) plots of Nu versus Pr. Also shown are the piecewise linear fits for Re and Nu with the breakpoints (exhibited by vertical dashed lines) computed using the ``pwlf" module~\cite{Jekel:PWLF2019} of python.}
	}
	\label{fig:ReNu_piecewise}
\end{figure}	
\begin{figure}[t]
	\includegraphics[scale=0.39]{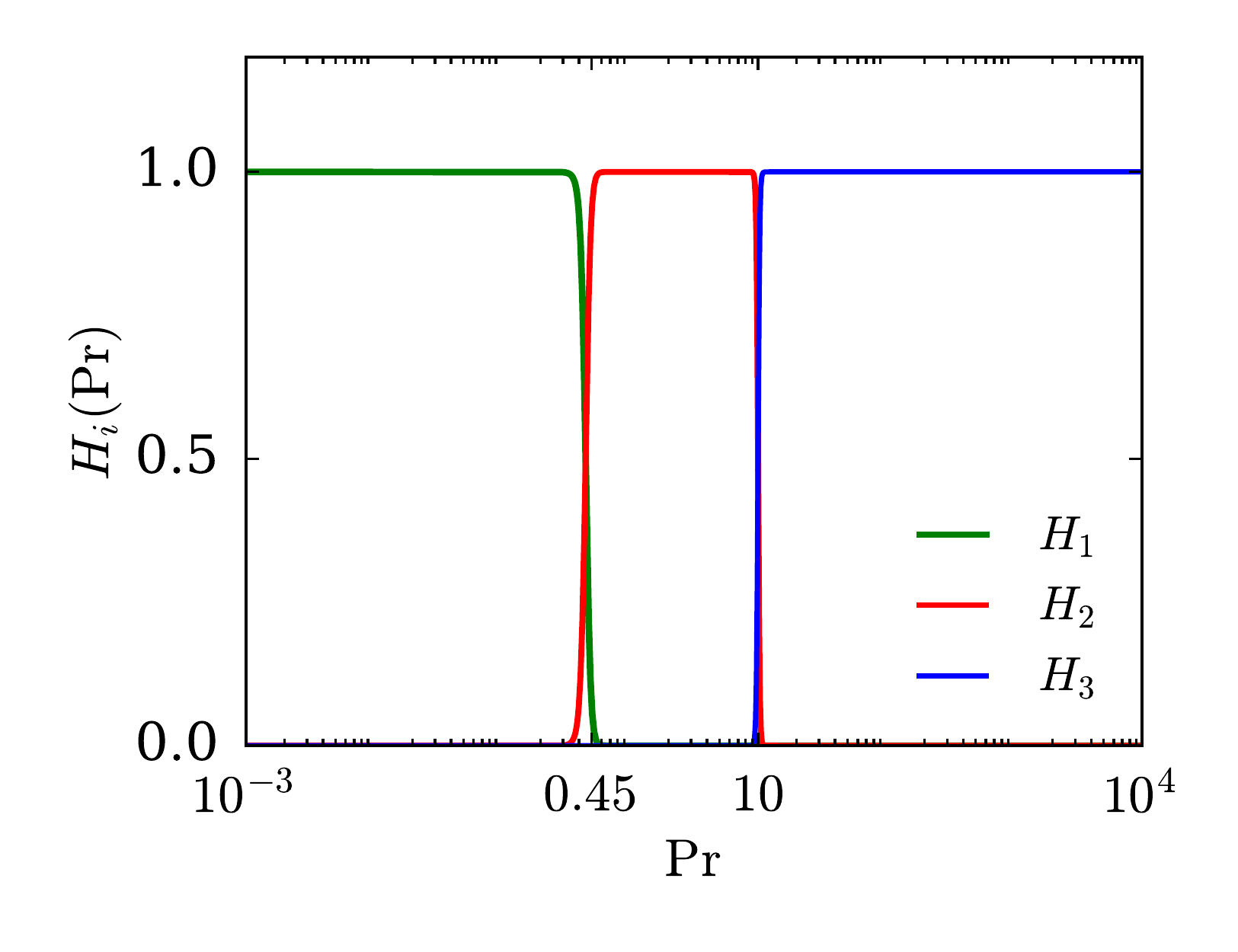}
	\caption{(color online) {\color{black}Plots of the matching functions $H_i(\mathrm{Pr})$ vs. Pr. $H_1$, $H_2$, and $H_3$ become unity in the regimes given by $\mathrm{Pr}<0.45$, $0.45<\mathrm{Pr}<10$, and $\mathrm{Pr}>10$ respectively. They attain the value of $1/2$ at the regime boundaries and become negligible outside their respective regimes.}}
	\label{fig:Match}
\end{figure}
\begin{table*}
	\caption{Details of the training dataset for the Multivariate Regression (MR) and the Neural Network (NN) models. The input parameters are the Prandtl number (Pr) and the Rayleigh number (Ra), and the output parameters are the Nusselt number (Nu) and the Reynolds number (Re).}
	\begin{ruledtabular}
		\begin{tabular}{c c c c c | c c c c c}
			S. No & Pr & Ra & Nu & Re & S. No & Pr & Ra & Nu & Re\\
			\hline 
			1 & 0.02 & $1 \times 10^5$ & 3.01 & 1180 & 32 & 6.8 & $1 \times 10^6$ & 7.90 & 24.9 \\
			2 & 0.02 & $2 \times 10^5$ & 3.52 & 1640 & 33 & 6.8 & $2 \times 10^6$ & 9.46 & 35.6 \\
			3 & 0.02 & $5 \times 10^5$ & 4.48 & 2440 & 34 & 6.8 & $5 \times 10^6$ & 12.9 & 59.7 \\
			4 & 0.02 & $1 \times 10^6$ & 5.78 & 3200 & 35 & 6.8 & $1 \times 10^7$ & 15.9 & 89.2 \\
			5 & 0.02 & $2 \times 10^6$ & 6.91 & 4290 & 36 & 6.8 & $2 \times 10^7$ & 19.5 & 128 \\
			6 & 0.02 & $5 \times 10^6$ & 8.85 & 6650 & 37 & 6.8 & $5 \times 10^7$ & 26.1 & 217 \\
			7 & 0.02 & $1 \times 10^7$ & 10.3 & 9420 & 38 & 6.8 & $1 \times 10^8$ & 31.6 & 314 \\
			8 & 0.1 & $5 \times 10^5$ & 6.11 & 749 & 39 & 6.8 & $2 \times 10^8$ & 38.5 & 452 \\
			9 & 0.1 & $1 \times 10^6$ & 7.34 & 1030 & 40 & 6.8 & $5 \times 10^8$ & 50.5 & 729 \\
			10 & 0.1 & $2 \times 10^6$ & 8.85 & 1380 & 41 & 6.8 & $1 \times 10^9$ & 65.7 & 1070 \\
			11 & 0.1 & $5 \times 10^6$ & 11.3 & 2090 & 42 & 6.8 & $2 \times 10^9$ & 77.0 & 1520 \\
			12 & 0.1 & $1 \times 10^7$ & 13.9 & 2870 & 43 & 6.8 & $5 \times 10^9$ & 101 & 2400 \\
			13 & 0.1 & $2 \times 10^7$ & 16.4 & 3870 & 44 & 50 & $1 \times 10^6$ & 8.17 & 3.53 \\
			14 & 0.1 & $5 \times 10^7$ & 20.8 & 6020 & 45 & 50 & $2 \times 10^6$ & 9.66 & 5.19 \\
			15 & 0.1 & $1 \times 10^8$ & 26.7 & 8140 & 46 & 50 & $5 \times 10^6$ & 13.8 & 9.38 \\
			16 & 0.5 & $1 \times 10^6$ & 8.38 & 285 & 47 & 50 & $1 \times 10^7$ & 16.7 & 14.0 \\
			17 & 0.5 & $3 \times 10^6$ & 11.4 & 482 & 48 & 50 & $2 \times 10^7$ & 20.2 & 21.1 \\
			18 & 0.5 & $1 \times 10^7$ & 15.9 & 874 & 49 & 50 & $5 \times 10^7$ & 26.4 & 35.2 \\
			19 & 0.5 & $3 \times 10^7$ & 21.6 & 1480 & 50 & 50 & $1 \times 10^8$ & 31.8 & 50.8 \\
			20 & 0.5 & $1 \times 10^8$ & 30.6 & 2610 & 51 & 50 & $2 \times 10^8$ & 38.7 & 76.4 \\
			21 & 1 & $1 \times 10^6$ & 8.18 & 147 & 52 & 50 & $5 \times 10^8$ & 51.8 & 137 \\
			22 & 1 & $2 \times 10^6$ & 10.1 & 213 & 53 & 50 & $1 \times 10^9$ & 61.5 & 202 \\
			23 & 1 & $5 \times 10^6$ & 13.3 & 340 & 54 & 100 & $1 \times 10^6$ & 7.94 & 1.80 \\
			24 & 1 & $1 \times 10^7$ & 16.3 & 491 & 55 & 100 & $2 \times 10^6$ & 10.4 & 2.78 \\
			25 & 1 & $2 \times 10^7$ & 19.8 & 702 & 56 & 100 & $5 \times 10^6$ & 13.9 & 4.90 \\
			26 & 1 & $5 \times 10^7$ & 26.0 & 1100 & 57 & 100 & $1 \times 10^7$ & 16.8 & 7.02 \\
			27 & 1 & $1 \times 10^8$ & 31.4 & 1530 & 58 & 100 & $2 \times 10^7$ & 20.1 & 9.91 \\
			28 & 1 & $2 \times 10^8$ & 38.6 & 2170 & 59 & 100 & $5 \times 10^7$ & 26.1 & 17.1 \\
			29 & 1 & $5 \times 10^8$ & 49.2 & 3330 & 60 & 100 & $1 \times 10^8$ & 31.8 & 26.0 \\
			30 & 1 & $1 \times 10^9$ & 61.2 & 4700 & 61 & 100 & $2 \times 10^8$ & 39.1 & 37.5 \\
			31 & 1 & $2 \times 10^9$ & 76.8 & 6580 & 62 & 100 & $5 \times 10^8$ & 49.7 & 71.4 \\   
		\end{tabular}
	\label{table:Training}
\end{ruledtabular}
\end{table*}

\begin{figure*}[t]
	\centering
	\includegraphics[scale=0.32]{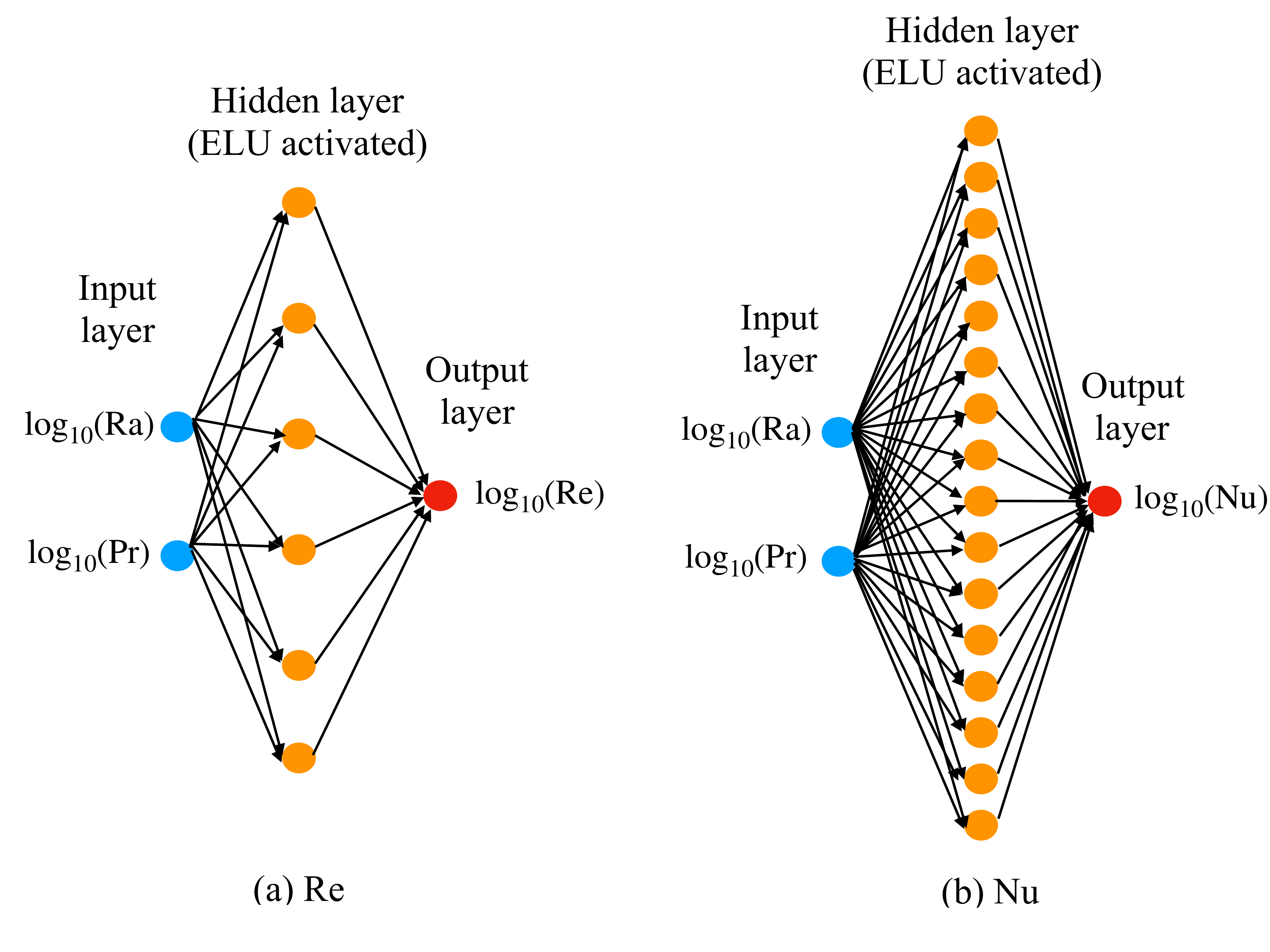}
	\caption{(color online) 
		A schematic diagram of the neural network models for the predictions of (a) Re and (b) Nu. Here, the input, hidden, and output layers are depicted by blue, orange and red nodes respectively. The models for Re and Nu have 6 and 16 hidden nodes respectively.}
	\label{fig:NN}
\end{figure*}
\subsection{Multivariate regression (MR) model}
\label{subsec:MR}
We construct multivariate regression (MR) model using machine-learning software WEKA~\cite{Weka:2009} and obtain the functional forms of Re and Nu. 
We look for a power-law relation for Re and Nu of the form $A\mathrm{Ra}^\alpha \mathrm{Pr}^\beta$, take logarithms of this expression, and employ linear regression to obtain $A$, $\alpha$, and $\beta$.  {\color{black}However, as discussed in Sec.~\ref{sec:Introduction}, the Pr and Ra dependences of Re and Nu are different at  small, moderate, and large Prandtl number regimes. Hence, the aforementioned power-law will not be accurate of all the regimes of Pr.

To overcome the above problem, we split our parameter space into three regimes -- small Pr, moderate Pr, and large Pr -- such that in each regime, the quantities $A$, $\alpha$, and $\beta$ do not vary significantly. We obtain the boundaries (or breakpoints) between the regimes as follows. For $\mathrm{Ra}=10^6$ and $10^7$, we plot the logarithms of Nu and Re versus the logarithm of Pr and fit a piecewise linear curve with three segments [see Figs.~\ref{fig:ReNu_piecewise}(a) and (b)]. We employ the ``PiecewiseLinFit" function from the ``pwlf" module of python~\cite{Jekel:PWLF2019}. The above module determines the breakpoints of the piecewise linear curve using an optimization process that solves a least squares fit several times for different breakpoint locations. The two breakpoints of Pr versus Nu and Pr versus Re curves determined by the ``pwlf" module for  $\mathrm{Ra}=10^6$ and $10^7$ are tabulated in Table~\ref{table:breakpoints}. Finally, we average the breakpoints over the two Rayleigh numbers and using this average, we obtain the parameter regimes for Re as
\begin{eqnarray}
\mbox{Small Pr} &:& \mathrm{Pr \leq 0.45}, \nonumber \\
\mbox{Moderate Pr} &:& 0.45 \leq \mathrm{Pr} \leq 10, \nonumber \\
\mbox{Large Pr} &:&  \mathrm{Pr} \geq 10, \nonumber
\end{eqnarray}
and the parameter regimes for Nu as
\begin{eqnarray}
\mbox{Small Pr} &:& \mathrm{Pr \leq 0.25}, \nonumber \\
\mbox{Moderate Pr} &:& 0.25 \leq \mathrm{Pr} \leq 4.6, \nonumber \\
\mbox{Large Pr} &:&  \mathrm{Pr} \geq 4.6. \nonumber
\end{eqnarray}
}

\begin{table}[t]
	{\color{black}
		\caption{For $\mathrm{Ra}=10^6$ and $10^7$: boundaries separating the small, moderate and large Pr regimes. Here, $\mathrm{Pr}_{l, \mathrm{Re}}$ = boundary between the small and moderate Pr regimes for the predictions of Re;  $\mathrm{Pr}_{u, \mathrm{Re}}$ = boundary between the moderate and large Pr regimes for the predictions of Re;  $\mathrm{Pr}_{l, \mathrm{Nu}}$ = boundary between the small and moderate Pr regimes for the predictions of Nu;  $\mathrm{Pr}_{u, \mathrm{Nu}}$ = boundary between the moderate and large Pr regimes for the predictions of Nu.}
		\label{table:breakpoints}
		\begin{center}
			\begin{tabular}{ccccc}
				\hline \hline
				Ra & $\mathrm{Pr}_{l, \mathrm{Re}}$ & $\mathrm{Pr}_{u,\mathrm{Re}}$ & $\mathrm{Pr}_{l, \mathrm{Nu}}$ & $\mathrm{Pr}_{u, \mathrm{Nu}}$  \\ \hline 
				$10^6$ & 0.25 & 6.8 & 0.28 & 2.5 \\ 
				$10^7$ & 0.65 & 14 & 0.22 & 6.8 \\ \hline
				Average & 0.45 & 10 & 0.25 & 4.6 \\ \hline
			\end{tabular}
		\end{center}
	}
\end{table}

To ensure  continuity across the parameter regimes, we employ {\color{black}six matching functions $H_i(\mathrm{Pr})$; $i=1$ to $6$. The functions $H_1$, $H_2$, $H_3$, $H_4$, $H_5$, and $H_6$ are such that they become unity inside the regimes defined by $\mathrm{Pr}<0.45$, $0.45<\mathrm{Pr}<10$,  $\mathrm{Pr}>10$, $\mathrm{Pr}<0.25$, $0.25<\mathrm{Pr}<4.6$, and $\mathrm{Pr}>4.6$, respectively, and become negligible outside their regimes. The value of these functions is 1/2 at the boundaries of their respective regimes. The following expressions of $H_i$ satisfy the aforementioned conditions: 
\begin{eqnarray}
H_1(\mathrm{Pr}) &=& \frac{1}{1+e^{-k_{1}(0.45-\mathrm{Pr})}}, \label{eq:H1} \\
H_2(\mathrm{Pr}) &=& \frac{1}{1+e^{-k_{1}(\mathrm{Pr}-0.45)}} - \frac{1}{1+e^{-k_{2}(\mathrm{Pr}-10)}}, \label{eq:H2} \\
H_3(\mathrm{Pr}) &=& \frac{1}{1+e^{-k_{2}(\mathrm{Pr}-10)}}, \label{eq:H3} 
\end{eqnarray}
\begin{eqnarray}
H_4(\mathrm{Pr}) &=& \frac{1}{1+e^{-k_{1}(0.25-\mathrm{Pr})}}, \label{eq:H4} \\
H_5(\mathrm{Pr}) &=& \frac{1}{1+e^{-k_{1}(\mathrm{Pr}-0.25)}} - \frac{1}{1+e^{-k_{2}(\mathrm{Pr}-4.6)}}, \label{eq:H5} \\
H_6(\mathrm{Pr}) &=& \frac{1}{1+e^{-k_{2}(\mathrm{Pr}-4.6)}}, \label{eq:H6}
\end{eqnarray}
where $k_1=50$ and $k_2=7$.  The above values of $k_1$ and $k_2$ provide the optimal steepness for the $H_i$ curves at the boundary points. See Fig.~\ref{fig:Match} for an illustration of the behavior of the matching functions $H_1$, $H_2$, and $H_3$.} We then determine the logarithms of the prefactor $A$ and the exponents $\alpha$ and $\beta$ for each regime using the \textit{linear regression} function of WEKA, and then combine these regimes using the matching functions $H_i$.

{\color{black}We make use of cross-validation technique to to get an insight on how the MR model will generalise to an independent dataset and thus to estimate how accurately it will perform in practice.}  We employ ten-fold cross-validation, which involves random partitioning of the complete dataset into 10 subdatasets of equal size. Of these 10 subdatasets, a single subdataset is retained as the validation data for testing the model, and the remaining 9 subdatasets are used as training data. This process is then repeated 10 times, with each of the 10 subdatasets used exactly once as the validation data. The mean of the absolute percentage error, given by 
\begin{equation}
\mathcal{D} = \left | \frac{\mbox{Predicted value} - \mbox{Actual value}}{\mbox{Actual value}} \right | \times 100,
\label{eq:Delta}
\end{equation}
is computed for all the samples of each fold and then averaged over all the folds. The above average is observed to be 4\% for the predictions of both Re and Nu, thus ensuring that the model is robust.
After having done the cross-validation, we invoke the learning algorithm for a final (11th) time on the entire dataset and obtain the following expressions for Re and Nu.
{\color{black}\begin{eqnarray}
\mathrm{Re} &=& 0.40 H_1 \mathrm{Ra}^{0.45} \mathrm{Pr}^{-0.72} + 0.11 H_2 \mathrm{Ra}^{0.52}\mathrm{Pr}^{-0.86} \nonumber \\
&&+ 0.055 H_3 \mathrm{Ra}^{0.58} \mathrm{Pr}^{-0.99}, \\
\mathrm{Nu} &=& 0.24 H_4 \mathrm{Ra}^{0.27} \mathrm{Pr}^{0.16} 
+  0.15 H_5 \mathrm{Ra}^{0.29} \mathrm{Pr}^{0.018}
\nonumber \\
&&+ 0.13 H_6 \mathrm{Ra}^{0.30} \mathrm{Pr}^{0.0084}. 
\end{eqnarray}}
{\color{black}Thus, interested researchers can use the above equations to obtain the Nusselt and Reynolds numbers for an arbitrary set of Rayleigh and Prandtl numbers.}

{\color{black}The above training algorithm was run on a personal computer consisting of an 8-gigahertz Core i5 processor with 8 gigabytes of RAM. The time taken for executing the training algorithm is 0.3 second for both Re and Nu.} It is important to note that the present model is an application of machine-learning model to the raw data itself, in contrast to RGL model where ML model was used for parameter estimation of the prefactors of the dissipation equations [Eqs.~(\ref{eq:TotalViscousDissipation}) and (\ref{eq:TotalThermalDissipation})].

\subsection{Neural Network (NN) model}
\label{subsec:NN}
Lastly, we  discuss  the neural network model. {\color{black}Neural network is known to be a more robust prediction algorithm than regression. 
This is due to the intrinsic nonlinearity of neural networks and also by the virtue of universal approximation theorems~\cite{Hornik:NN1989} associated with these networks.} A schematic diagram of the neural networks for predicting Re and Nu are exhibited in Fig.~\ref{fig:NN}(a,b). We employ the Keras module~\cite{Gulli:book} of Python to construct the neural networks. The input features for both the networks are the logarithms of Rayleigh and Prandtl numbers. The networks consist of one hidden layer, with 6 nodes for Re predictions and 16 nodes for Nu predictions. We arrived at the above configuration after trial and error as it provided us the best results. {\color{black} It must be noted from our discussion in Sec.~\ref{sec:Introduction} that the range of the scaling exponent for Nu ($0.25 \leq \alpha \leq 0.33$) is shorter than that for Re ($0.40 \leq \gamma \leq 0.60$); hence, more care is needed for predicting Nu. Thus, more neurons are required in the neural network for Nu.}

The hidden layers of both the networks are activated by the \textit{Exponential Linear Unit} (ELU) function~\cite{Goodfellow:book}. The networks are set to minimize the mean-squared error between the actual values in the training set and the predictions. We employ the \textit{Adam} algorithm~\cite{Gulli:book} to optimize our models. The networks are trained using the \textit{mini-batch} gradient descent algorithm with a batch size of 8, that is, eight training examples are used to estimate the error gradient. The gradient descent algorithm is run for 1000 epochs.  {\color{black}The training algorithms were run on the same personal computer on which the MR algorithm was executed. The computational time taken to train the model is 11 seconds for both Re and Nu; thus the neural networks take more time than the multivariate regression for training. However, the time required for training is still not very large and the necessary computations can easily be carried out on a laptop.} The relevant segment and functions of the code describing the model (for Re) are given below.\\ \\ 
\texttt{def baseline\_model():} \\
\indent \texttt{model = Sequential()} \\
\indent \textit{'\thinspace'\thinspace' Specify the number of input parameters (input\_dim), \\
	\indent which in this case is 2. Add the hidden layer with \\
	\indent 6 nodes, activated using ELU function.'\thinspace'\thinspace'} \\
\indent \texttt{model.add(Dense(6, input\_dim=2, \\
	\indent kernel\_initializer = 'normal', activation = 
	\indent 'elu'))} \\
\indent \textit{\# Add the output layer containing one output node.} \\ 
\indent \texttt{model.add(Dense(1, kernel\_initializer = \\
	\indent 'normal'))} \\
\indent \textit{\# Specify the loss function and the optimizer} \\
\indent \texttt{model.compile(loss='mean\_squared\_error', \\
	\indent optimizer='adam')} \\
\indent \texttt{return model} \\ 
\textit{\# Specify the batch size and the number of epochs} \\
\texttt{estimator = KerasRegressor(build\_fn = baseline\_model, epochs = 1000, batch\_size = 8, verbose = 0)} \\ 
\textit{\#Functions for training and predicting:} \\ 
\textit{\#Train the model (X = input, Y = output)} \\
\texttt{estimator.fit(X,Y)} \\
\textit{\#Predict the values for all X} \\
\texttt{predictions = estimator.predict(X)}
\\

The relevant code segment for describing the model for Nu is similar to that for Re, except that the number of nodes in the hidden layer is specified to be 16. {\color{black} The user can pass a two-dimensional array of arbitrary Rayleigh and Prandtl numbers as an argument to the ``estimator.predict()" function; this function returns the corresponding array for predictions.} Refer to the supplemental materials for the complete codes (``suppl1.txt" for predicting Re and ``suppl2.txt" for predicting Nu) and the input dataset (``suppl3.csv"). 
	
We employ 10-fold cross-validation for our above models (similar to that done for the MR model described earlier) to get an estimate of how the models generalize to an independent dataset. We observe the mean absolute error from the cross-validation to be 4\% for the predictions of Re and 3\% for the predictions of Nu. The small values of the errors indicate that our neural-network model has a robust predictive capability.
{\color{black}It must be noted, however, that cross-validation may be faulty in determining the prediction reliability of a machine-learning model in certain cases~\cite{Bishop:book:2006} and a more appropriate tool is to compare the absolute percentage error between the final model (developed using entire training dataset) and the actual values of the test datasets. This will be covered in detail in Sec.~\ref{sec:Results}.}

 {\color{black}To ascertain the absence of overfitting, we add weighted regularizing terms (of $\mathrm{L}^2$ norm~\cite{Goodfellow:book}) to our neural network models and observe how the models fair against the test data. In the present work, the test data consists of the simulation results of \citet{Scheel:PRF2017} ($\mathrm{Pr} =0.005$ and $0.02$), \citet{Wagner:PF2013} ($\mathrm{Pr}=0.7$), \citet{Emran:JFM2008} ($\mathrm{Pr}=0.7$),  \citet{Kaczorowski:JFM2013} ($\mathrm{Pr}=4.38$), and \citet{Horn:JFM2013} ($\mathrm{Pr}=2547.9$); and the experimental results of \citet{Cioni:JFM1997} ($\mathrm{Pr}=0.02$) and \citet{Niemela:JFM2001} ($\mathrm{Pr}=0.7$). 
\begin{table}[t]
	{\color{black}
		\caption{Effect of adding regularizing hyperparameters to the neural networks on the performance of the models on test data. Here, $D_\mathrm{Re}$ and $D_\mathrm{Nu}$ respectively represent the percentage deviation of the predictions from the actual values of the test data.}
		\label{table:Regularization}
		\begin{center}
			\begin{tabular}{ccc}
				\hline \hline
				Regularizer hyperparameter & $D_\mathrm{Re}$ & $D_\mathrm{Nu}$ \\ \hline \\
				None & 19.2\% & 9.76\%\\ \\
				0.001 (Hidden layer) & 20.4\% & 14.0\%\\ \\
				0.001 (Hidden layer) & 20.5\% & 14.4\%\\
				0.001 (Output layer) & & \\ \\
				0.01 (Hidden layer) & 21.7\% & 25.8\%\\ \\
				0.01 (Hidden layer) & 23.5\% & 26.1\%\\
				0.01 (Output layer) & & \\ \hline
			\end{tabular}
		\end{center}
	}
\end{table}
 In Table~\ref{table:Regularization}, we list the regularizing hyperparameters added in the hidden and the output layers of the neural networks and their effects on the accuracy of the networks. The table shows that the addition of regularizing terms does not improve the forecast; rather the forecast worsens as the values of the  hyperparameters increase. Thus, in the final neural network models, no regularizing counterterms are used.}

\begin{table*}[t]
	{\color{black}
		\caption{Different trial configurations of neural networks and their corresponding performances. Here, HL = hidden layer, $D_\mathrm{Re}$ = percentage deviation of the predictions from the actual Re, $D_\mathrm{Nu}$ = percentage deviation of the predictions from the actual Nu.}
		\label{table:NN_config}
		\begin{center}
			\begin{tabular}{cccccc}
				\hline \hline
				Configuration & $D_\mathrm{Re}$ & $D_\mathrm{Re}$ & Configuration & $D_\mathrm{Nu}$  & $D_\mathrm{Nu}$ \\ 
				(Re) & (training) & (testing) & (Nu) & (training) & (testing) \\ \hline \\
				1 HL, & & & 1 HL, & & \\
				6 nodes per HL & 2.98\% & 19.2\% & 16 nodes per HL & 2.31\% & 9.76\% \\  \\
				2 HL, & & & 2 HL, & & \\
				3 nodes per HL & 2.98\% & 36.0\%& 8 nodes per HL & 3.04\% & 17.9\% \\ \\
				3 HL, & &  & 4 HL & & \\ 
				2 nodes per HL & 5.93\% & 81.6\% & 4 nodes per HL & 4.64\% & 17.1\% \\ \hline
			\end{tabular}
		\end{center}
	}
\end{table*}
{\color{black}Finally, we remark that we also tried to use less width and more depth to our neural networks. As shown in Table~\ref{table:NN_config}, we examined the performance of the neural network using two and three hidden layers for predicting Re and two and four hidden layers for predicting Nu. In all the above trials, we kept the same total number of hidden nodes (that is, 6 for Re and 16 for Nu). Interestingly, we observed that the performance of the networks decreases with the increase in depth. This is likely because the deeper neural networks have a tendency to overfit and hence their predictions are less accurate.}

 The important aspects of the MR and NN models are summarized in Table~\ref{table:Models}. In the next section, we discuss and compare the predictions of these models with respect to the GL, RGL, and PV models.
\section{Performance of the ML models}
\label{sec:Results}
\begin{table}[t]
	\caption{Absolute percentage error in the predictions of Re by the GL model, the RGL model, the PV model, the multivariate regression (MR) model, and the neural network (NN) model for different sets of simulation and experimental data. The asterisks (*) refer to the numerical data of \citet{Bhattacharya:PF2021}, used for training the MR and NN models.}
	\label{table:Comparison_Re}
	\begin{center}
		\begin{tabular}{cccccc}
			\hline \hline
			Pr & GL & RGL & PV & MR & NN\\ \hline
			0.005 & 48\% & 11\% & 11\% & {\color{black}13\%} & 4.5\%\\
			$0.02^*$ & 52\% & 4.7\% & 31\% & {\color{black}3.5\%} & {\color{black}8.0\%}  \\
			0.02 & 53\% & 11\% & 19\% & {\color{black}8.8\%} & 9.0\%  \\
			$0.1^*$ & 30\% & 1.9\% & 11\% & {\color{black}3.2\%} & 3.4\% \\
			$0.5^*$ & 14\% & 1.3\% & 3.3\% & {\color{black}4.5\%} & 3.1\%  \\
			0.7 & 24\% & 6.7\% & 23\% & {\color{black}13\%} & 9.2\% \\
			$1^*$ & 20\% & 2.8\% & 8.6\% & {\color{black}5.1\%} & 1.4\% \\
			$6.8^*$ & 27\% & 3.4\% & 8.4\% & {\color{black}5.2\%} & 2.8\% \\
			$50^*$ & 84\% & 6.0\% & 12\% & {\color{black}4.5\%} & 3.4\%  \\
			$100^*$ & 150\% & 3.8\% & 15\% & {\color{black}4.6\%} & 3.6\% \\
			2547.9 & 560\% & 85\% & 77\% & 36\% & 37\% \\  \hline
			Overall & 82\% & 12\% & 20\% & {\color{black}9.3\%} & 7.7\% \\ \hline \hline
		\end{tabular}
	\end{center}
\end{table}

\begin{table}[htbp]
	\caption{Absolute percentage error in the predictions of Nu by the GL model, the RGL model, the PV model, the multivariate regression (MR) model, and the neural network (NN) model for different sets of simulation and experimental data. The asterisks (*) refer to the numerical data of \citet{Bhattacharya:PF2021}, used for training the MR and NN models.}
	\label{table:Comparison_Nu}
	\begin{center}
		\begin{tabular}{cccccc}
			\hline \hline
			Pr  & GL & RGL & PV & MR & NN \\ 
			\hline
			0.005 & 17\% & 9.6\% & 93\% & {\color{black}3.3\%} & 2.7\% \\
			$0.02^*$ & 34\% & 8.9\% & 150\% & {\color{black}2.2\%} & 5.1\% \\
			0.02 & 15\% & 11\% & 84\% & {\color{black}6.1\%} & 12\% \\
			$0.1^*$ & 5.0\% & 3.1\% & 22\% & {\color{black}6.2\%} & 2.5\% \\
			$0.5^*$ & 5.4\% & 1.4\% & 6.0\% & {\color{black}1.5\%} & 1.6\% \\
			0.7 & 11\% & 9.1\% & 9.7\% & 6.5\% & 3.0\% \\
			$1^*$ & 5.8\% & 3.6\% & 9.2\% & {\color{black}1.1\%} & 0.85\% \\
			4.38 & 6.3\% & 5.7\% & 7.6\% & {\color{black}1.0\%} & 0.95\% \\
			$6.8^*$ & 6.4\% & 7.2\% & 3.8\% & {\color{black}5.1\%} & 3.3\% \\
			$50^*$ & 7.2\% & 3.2\% & 7.3\% & {\color{black}4.9\%} & 1.9\% \\
			$100^*$ & 3.9\% & 2.7\% & 6.4\% & {\color{black}4.9\%} & 2.1\% \\
			2547.9 & 17\% & 2.3\% & 7.6\% & {\color{black}4.7\%} & 14\% \\ \hline
			Overall & 11\% & 5.7\% & 34\% & {\color{black}4.0\%} & 4.2\% \\
			\hline \hline
		\end{tabular}
	\end{center}
\end{table}

\begin{figure*}[t]
	\centering
	\includegraphics[scale=0.4]{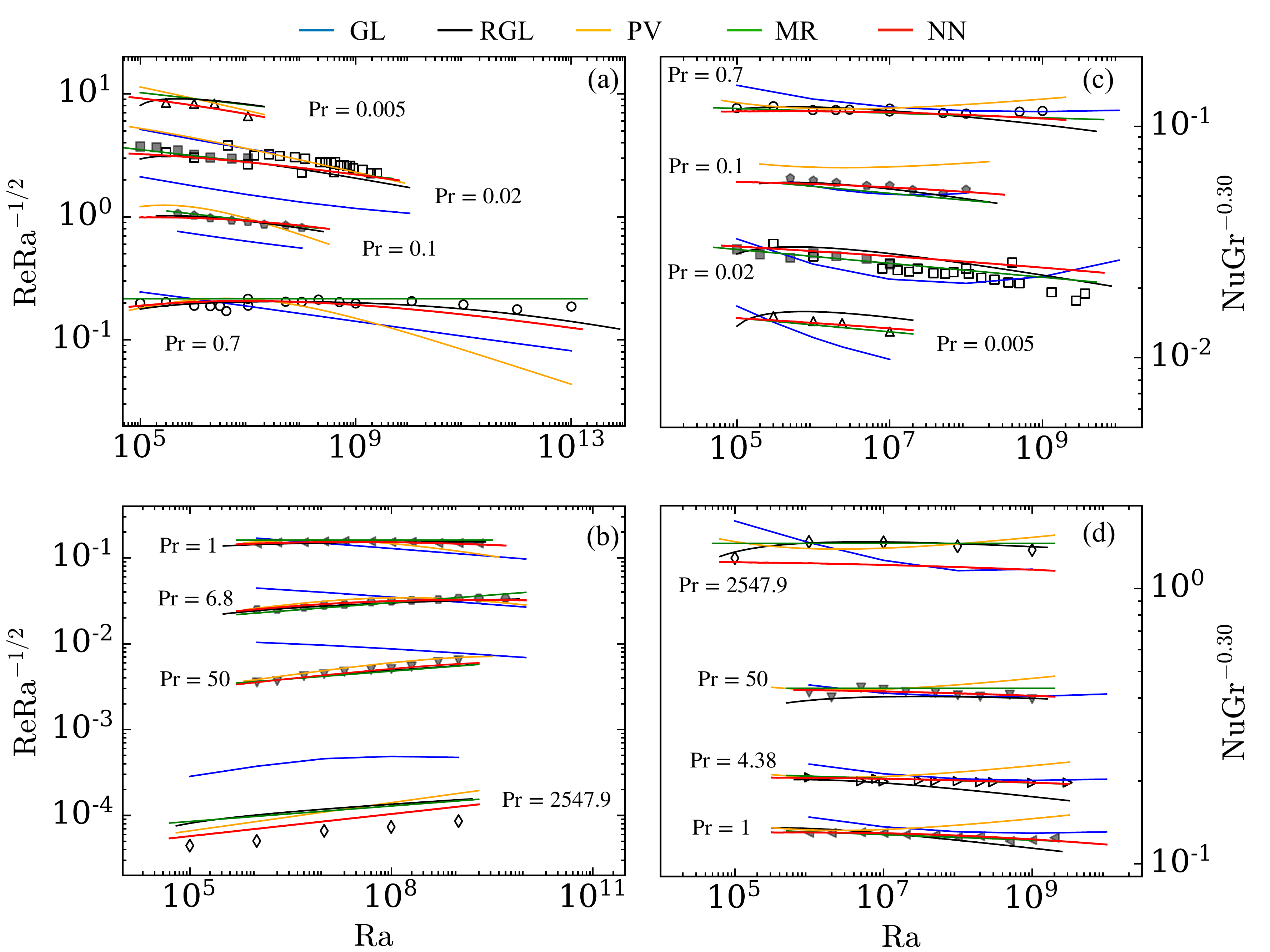}
	\caption{(color online) 
		Comparison between Grossmann and Lohse (GL, blue lines), Pandey and Verma (PV, orange lines), Revised Grossmann and Lohse (RGL, black lines), multivariate regression (MR, green lines), and the neural network (NN, red lines) models for the predictions of Re for (a) $\mathrm{Pr} < 1$, and (b) $\mathrm{Pr}\geq 1$; and of Nu for (c) $\mathrm{Pr} < 1$, and (d) $\mathrm{Pr}\geq 1$. 
		The predictions are compared with the past numerical and experimental results for $\mathrm{Pr}=0.005$ (triangles), $\mathrm{Pr}=0.02$ (squares), $\mathrm{Pr}=0.1$ (pentagons), $\mathrm{Pr}=0.7$ (circles), $\mathrm{Pr}=1$ (left-pointing triangles), $\mathrm{Pr}=4.38$ (right-pointing triangles), $\mathrm{Pr}=6.8$ (hexagons), $\mathrm{Pr}=50$ (inverted triangles), and $\mathrm{Pr}=2547.9$ (diamonds). 
		The filled markers correspond to the training dataset for the MR and NN models that includes the numerically computed values of \citet{Bhattacharya:PRF2021}.}
	\label{fig:Re_Nu}
\end{figure*}

We  test the performance of the multivariate regression and neural network models  with the GL, RGL, and PV models.  We  compare their predictions for the numerical results of \citet{Bhattacharya:PF2021}($\mathrm{Pr}=0.02$, 0.1, 0.5, 1, 6.8, 50, and 100), \citet{Scheel:PRF2017} ($\mathrm{Pr} =0.005$ and $0.02$), along with the test dataset \citet{Wagner:PF2013} ($\mathrm{Pr}=0.7$), \citet{Emran:JFM2008} ($\mathrm{Pr}=0.7$),  \citet{Kaczorowski:JFM2013} ($\mathrm{Pr}=4.38$), and \citet{Horn:JFM2013} ($\mathrm{Pr}=2547.9$); and the experimental results of \citet{Cioni:JFM1997} ($\mathrm{Pr}=0.02$), and \citet{Niemela:JFM2001} ($\mathrm{Pr}=0.7$). The simulations of \citet{Wagner:PF2013}, \citet{Kaczorowski:JFM2013}, and \citet{Bhattacharya:PF2021} involve a cubical cell, whereas the others employ a cylindrical cell. The aspect ratio of the RBC cell is unity in all the above works.
We compute the absolute percentage error between the estimated and actual values {\color{black}as per Eq.~(\ref{eq:Delta}).
For every Pr, we average the aforementioned errors over all the Rayleigh numbers and list this average error in Table~\ref{table:Comparison_Re} (for the predictions of Re) and Table~\ref{table:Comparison_Nu} (for the predictions of Nu).} Note that for $\mathrm{Pr}=0.02$, we provide two entries in the above tables: one corresponding to the data of \citet{Bhattacharya:PF2021} (which were used for training), and the other corresponding to those of \citet{Scheel:PRF2017} and \citet{Cioni:JFM1997}. In Fig.~\ref{fig:Re_Nu}(a,b), we plot the  predictions of the normalized Re by the different models discussed in the paper along with the numerically computed values of Refs.~\cite{Bhattacharya:PF2021,Scheel:PRF2017,Wagner:PF2013,Emran:JFM2008,Horn:JFM2013}. See Fig.~\ref{fig:Re_Nu}(c, d) for the corresponding plots for Nu. Note that in the above figures, we normalize Nu with $\mathrm{Gr}^{0.3}$, where $\mathrm{Gr}=\mathrm{Ra}/\mathrm{Pr}$ is the Grashoff number. This is because Nu has only a weak dependence on Pr, due to which the Nu versus Ra curves for different Pr's would have tended to overlap on each other. Further, we do not show the plots for $\mathrm{Pr}=6.8$ (only for Nu) and $\mathrm{Pr}=0.5$ in the above figures because their datapoints lie very close to those of $\mathrm{Pr}=4.38$ and $0.7$ respectively.

\subsection{Predictions of the Reynolds number}
\label{subsec:Re_predictions}
First, we analyse in detail the performance of the models in their predictions of the Reynolds number. From Table~\ref{table:Comparison_Re} and Figs.~\ref{fig:Re_Nu}(a,b), we observe that the GL model's predictions  for Re are close to the actual values for moderate Prandtl numbers. Note, however, that Re is sensitive to the way it is computed (that is, based on root mean square velocity, maximum plume velocity, etc.). The RGL model further improves the predictions of Re, especially for extreme Prandtl numbers. The PV model also marginally improves the predictions of the Reynolds number compared to the GL model but is not as accurate as the RGL model. 

{\color{black}Now, we will analyze the performances of the MR and NN models developed in the present study. Both the above models exhibit very good accuracy in their predictions of Re and provide significantly better predictions than the GL and the PV models. As exhibited in Table~\ref{table:Comparison_Re}, the MR and NN models provide better predictions than even the RGL model for large Prandtl numbers. For small Pr's, however, the MR model fares marginally worse than the NN and RGL models, with the accuracy of the latter two models being comparable.  It must be noted, however, that the MR model performs better than the RGL model if we consider all the Prandtl numbers into account (as shown in Table~\ref{table:Comparison_Re}).}


\subsection{Predictions of the Nusselt number}
\label{subsec:Nu_predictions}
Having analyzed the predictions of Re, we will now examine the preformance of the models in predicting Nu. 
Figures.~\ref{fig:Re_Nu}(c,d) and Table~\ref{table:Comparison_Nu} show that the GL model predicts Nu with a good accuracy. The RGL model marginally improves the predictions of Nu for moderate Pr but significantly improves the predictions for extreme Pr. 
 The PV model's predictions of Nu are close to those by the GL model for moderate and large Pr. However, for small Prandtl numbers, Nu predicted by the PV model deviates significantly from the actual values.  
This is because  Pandey and Verma~\cite{Pandey:PF2016,Pandey:PRE2016} derived their expression for Nu  using the data from simulations with $\mathrm{Pr} \geq 1$. As a result, the final expression for the Nusselt number did not contain any Pr dependence, which is applicable only for moderate and large Prandtl numbers. Hence, for small Prandtl numbers, the errors between the predictions and the actual Nu values are very high for the PV model. {\color{black}Thus, as far as the previously developed models are concerned, the RGL model exhibits the best performance.}

{\color{black} The MR and NN models developed in this study exhibit very good accuracy in their predictions of Nu and are observed to mostly perform better than the GL and PV models. The predictions of the MR and NN models are even better, albeit marginally, than the RGL model. For small and moderate Pr's ($\mathrm{Pr} \lesssim 1$), both MR and NN models perform better than the RGL model, with the MR model marginally outperforming the NN model. For larger Pr's, with the exception of $\mathrm{Pr} < 2547.9$, the NN model performs better than the MR and RGL models. The performances of the latter two models are comparable in the aforementioned regime. For $\mathrm{Pr} = 2547.9$, the ML models, especially the NN model, do not perform as well as they do for the remaining Pr's; this is possibly because of lack of data points for very large Pr's.}

\subsection{Significance of the ML models}
\label{subsec:Significance}
{\color{black}In  Sec.~\ref{subsec:Re_predictions} and Sec.~\ref{subsec:Nu_predictions}, we have shown that the ML models developed in this study exhibit better accuracy in predictions compared to GL and PV models, Further, although the ML models offer better predictions than the RGL model as well, the improvement is only marginal. In spite of the above, the ML models are more promising than the RGL model as they easier to construct provided we have sufficient  training dataset. Fortunately, there is a large amount of datasets available in the literature (see, for example, \citet{Ahlers:RMP2009}, \citet{Chilla:EPJE2012}).  Also, experiments and numerical simulations of the past provide a wide  range of Ra and Pr for training the neural network.  For example, \citet{Ahlers:NJP2012}, \citet{He:PRL2012}, and \citet{Urban:PRL2012} performed experiments for Ra as large as $10^{15}$, while  \citet{Niemela:Nature2000} reached Ra up to $10^{17}$.  In addition, \citet{Zhu:PRL2018} and \citet{Iyer:PNAS2020} recently carried out numerical simulations for Ra  up to $10^{14}$ and $10^{15}$ respectively. Thus, the machine learning models developed in this study has a good scope for improvement with inclusion of these datasets. 
 	
On the other hand, the training dataset for the RGL model is more difficult to construct because it requires the prefactors for the viscous and thermal dissipation rates in the bulk and boundary layers as output parameters. The data on these prefactors is quite limited; in fact, to the best of our knowledge, only \citet{Bhattacharya:PF2021} have computed the above quantities using numerical data of RBC. Further, there is no available data on $f_i$'s for extreme Rayleigh numbers ($\gg 10^{10}$). Note that most of the work on such large Rayleigh numbers are experimental because simulations of RBC in that regime require large computational resources. However, computation of prefactors for the dissipation rates requires multipoint measurements, which is  difficult  in experiments.  Due to these constraints of the RGL model, we believe that NN and MR models hold a lot of promise for a robust model for forecasting Re and Nu. }
We conclude in the next section.

\section{Summary and conclusions}
\label{sec:Conclusion}
In summary, we construct two ML models--multivariate regression (MR) model and neural network (NN) model --to predict Re and Nu for given Ra and Pr.  We compared the models' predictions with existing models.   We observe that although all the above models predict Re and Nu with good accuracy, {\color{black}the ML models exhibit the best performace.} In general, we observed that Re is more sensitive to modeling parameters compared to Nu; this is because  Re depends more critically on Pr compared to Nu. We also discussed that the {\color{black}machine learning models developed in this study} can be enhanced further by using data for a wider range of parameters.  

The positive results from the {\color{black}machine learning models} reinforce the importance of  data-driven methods for modelling turbulence, a practice that is gaining popularity~\cite{Pandey:PRF2020,Pandey:JOT2020,Fonda:PNAS2019,Parish:JCP2016}.   {\color{black}An accurate estimation of the large-scale velocity and heat transport helps scientists and engineers to better understand thermally driven flows encountered in nature and in engineering applications. For example, our results will be useful in understanding atmospheric flows and thus can aid in more accurate climate modelling and weather predictions. On the other hand, our findings will help engineers to better design buildings, heat sinks, solar collectors, and chimneys. Further, since the improvement in the predictions by our models is more pronounced for extreme Prandtl numbers, we believe that our results will be useful in modelling natural flows and designing applications where such extreme Prandtl numbers are involved. Examples of above include the mantle flow (very large Prandtl number) and flows in liquid metal batteries (small Prandtl number).}
We hope that in future, more machine-learning tools will be constructed for thermal convection that will help model natural and engineering flows.

\section*{Acknowledgements}
We thank J. Schumacher, K. R. Sreenivasan, A. Pandey, M. Sharma, R. Samuel, and S. Alam for useful discussions. The present work was mostly conducted at Indian Institute of Technology Kanpur, India, and was funded by the research grant number SPO/STC/PHY/2018037 from Indian Space Research Organization, India. The simulations of convection were performed on Shaheen II of King Abdullah University of Science and Technology, Saudi Arabia, under the project k1416. Shashwat Bhattacharya is currently funded by a postdoctoral fellowship of Alexander von Humboldt Foundation (Germany).

\section*{Conflicts of interest}
The authors have no conflicts to disclose.

\section*{Supplementary Material}
See supplementary material for the python codes for predicting Re (suppl1.txt) and Nu (suppl2.txt) using neural network, as well as the training dataset (suppl3.csv).

\section*{References}

\begin{thebibliography}{69}%
	\makeatletter
	\providecommand \@ifxundefined [1]{%
		\@ifx{#1\undefined}
	}%
	\providecommand \@ifnum [1]{%
		\ifnum #1\expandafter \@firstoftwo
		\else \expandafter \@secondoftwo
		\fi
	}%
	\providecommand \@ifx [1]{%
		\ifx #1\expandafter \@firstoftwo
		\else \expandafter \@secondoftwo
		\fi
	}%
	\providecommand \natexlab [1]{#1}%
	\providecommand \enquote  [1]{``#1''}%
	\providecommand \bibnamefont  [1]{#1}%
	\providecommand \bibfnamefont [1]{#1}%
	\providecommand \citenamefont [1]{#1}%
	\providecommand \href@noop [0]{\@secondoftwo}%
	\providecommand \href [0]{\begingroup \@sanitize@url \@href}%
	\providecommand \@href[1]{\@@startlink{#1}\@@href}%
	\providecommand \@@href[1]{\endgroup#1\@@endlink}%
	\providecommand \@sanitize@url [0]{\catcode `\\12\catcode `\$12\catcode
		`\&12\catcode `\#12\catcode `\^12\catcode `\_12\catcode `\%12\relax}%
	\providecommand \@@startlink[1]{}%
	\providecommand \@@endlink[0]{}%
	\providecommand \url  [0]{\begingroup\@sanitize@url \@url }%
	\providecommand \@url [1]{\endgroup\@href {#1}{\urlprefix }}%
	\providecommand \urlprefix  [0]{URL }%
	\providecommand \Eprint [0]{\href }%
	\providecommand \doibase [0]{http://dx.doi.org/}%
	\providecommand \selectlanguage [0]{\@gobble}%
	\providecommand \bibinfo  [0]{\@secondoftwo}%
	\providecommand \bibfield  [0]{\@secondoftwo}%
	\providecommand \translation [1]{[#1]}%
	\providecommand \BibitemOpen [0]{}%
	\providecommand \bibitemStop [0]{}%
	\providecommand \bibitemNoStop [0]{.\EOS\space}%
	\providecommand \EOS [0]{\spacefactor3000\relax}%
	\providecommand \BibitemShut  [1]{\csname bibitem#1\endcsname}%
	\let\auto@bib@innerbib\@empty
	\bibitem [{\citenamefont
		{Chandrasekhar}(1981)}]{Chandrasekhar:book:Instability}%
	\BibitemOpen
	\bibfield  {author} {\bibinfo {author} {\bibfnamefont {S.}~\bibnamefont
			{Chandrasekhar}},\ }\href@noop {} {\emph {\bibinfo {title} {{Hydrodynamic and
					Hydromagnetic Stability}}}}\ (\bibinfo  {publisher} {Dover publications},\
	\bibinfo {address} {Oxford},\ \bibinfo {year} {1981})\BibitemShut {NoStop}%
	\bibitem [{\citenamefont {Ahlers}, \citenamefont {Grossmann},\ and\
		\citenamefont {Lohse}(2009)}]{Ahlers:RMP2009}%
	\BibitemOpen
	\bibfield  {author} {\bibinfo {author} {\bibfnamefont {G.}~\bibnamefont
			{Ahlers}}, \bibinfo {author} {\bibfnamefont {S.}~\bibnamefont {Grossmann}}, \
		and\ \bibinfo {author} {\bibfnamefont {D.}~\bibnamefont {Lohse}},\ }\bibfield
	{title} {\enquote {\bibinfo {title} {{Heat transfer and large scale dynamics
					in turbulent Rayleigh-B{\'e}nard convection}},}\ }\href@noop {} {\bibfield
		{journal} {\bibinfo  {journal} {Rev. Mod. Phys.}\ }\textbf {\bibinfo {volume}
			{81}},\ \bibinfo {pages} {503--537} (\bibinfo {year} {2009})}\BibitemShut
	{NoStop}%
	\bibitem [{\citenamefont {Chill{\`a}}\ and\ \citenamefont
		{Schumacher}(2012)}]{Chilla:EPJE2012}%
	\BibitemOpen
	\bibfield  {author} {\bibinfo {author} {\bibfnamefont {F.}~\bibnamefont
			{Chill{\`a}}}\ and\ \bibinfo {author} {\bibfnamefont {J.}~\bibnamefont
			{Schumacher}},\ }\bibfield  {title} {\enquote {\bibinfo {title} {{New
					perspectives in turbulent Rayleigh-B{\'e}nard convection}},}\ }\href@noop {}
	{\bibfield  {journal} {\bibinfo  {journal} {Eur. Phys. J. E}\ }\textbf
		{\bibinfo {volume} {35}},\ \bibinfo {pages} {58} (\bibinfo {year}
		{2012})}\BibitemShut {NoStop}%
	\bibitem [{\citenamefont {Siggia}(1994)}]{Siggia:ARFM1994}%
	\BibitemOpen
	\bibfield  {author} {\bibinfo {author} {\bibfnamefont {E.~D.}\ \bibnamefont
			{Siggia}},\ }\bibfield  {title} {\enquote {\bibinfo {title} {{High Rayleigh
					number convection}},}\ }\href@noop {} {\bibfield  {journal} {\bibinfo
			{journal} {Annu. Rev. Fluid Mech.}\ }\textbf {\bibinfo {volume} {26}},\
		\bibinfo {pages} {137--168} (\bibinfo {year} {1994})}\BibitemShut {NoStop}%
	\bibitem [{\citenamefont {Xia}(2013)}]{Xia:TAML2013}%
	\BibitemOpen
	\bibfield  {author} {\bibinfo {author} {\bibfnamefont {K.-Q.}\ \bibnamefont
			{Xia}},\ }\bibfield  {title} {\enquote {\bibinfo {title} {{Current trends and
					future directions in turbulent thermal convection}},}\ }\href@noop {}
	{\bibfield  {journal} {\bibinfo  {journal} {Theor. App. Mech. Lett.}\
		}\textbf {\bibinfo {volume} {3}},\ \bibinfo {pages} {052001} (\bibinfo {year}
		{2013})}\BibitemShut {NoStop}%
	\bibitem [{\citenamefont {Verma}(2018)}]{Verma:book:BDF}%
	\BibitemOpen
	\bibfield  {author} {\bibinfo {author} {\bibfnamefont {M.~K.}\ \bibnamefont
			{Verma}},\ }\href@noop {} {\emph {\bibinfo {title} {Physics of Buoyant Flows:
				From Instabilities to Turbulence}}}\ (\bibinfo  {publisher} {World
		Scientific},\ \bibinfo {address} {Singapore},\ \bibinfo {year}
	{2018})\BibitemShut {NoStop}%
	\bibitem [{\citenamefont {Malkus}(1954)}]{Malkus:PRSA1954}%
	\BibitemOpen
	\bibfield  {author} {\bibinfo {author} {\bibfnamefont {W.~V.~R.}\
			\bibnamefont {Malkus}},\ }\bibfield  {title} {\enquote {\bibinfo {title}
			{{The Heat Transport and Spectrum of Thermal Turbulence}},}\ }\href@noop {}
	{\bibfield  {journal} {\bibinfo  {journal} {Proceedings of the Royal Society
				of London. Series A}\ }\textbf {\bibinfo {volume} {225}},\ \bibinfo {pages}
		{196--212} (\bibinfo {year} {1954})}\BibitemShut {NoStop}%
	\bibitem [{\citenamefont {Shraiman}\ and\ \citenamefont
		{Siggia}(1990)}]{Shraiman:PRA1990}%
	\BibitemOpen
	\bibfield  {author} {\bibinfo {author} {\bibfnamefont {B.~I.}\ \bibnamefont
			{Shraiman}}\ and\ \bibinfo {author} {\bibfnamefont {E.~D.}\ \bibnamefont
			{Siggia}},\ }\bibfield  {title} {\enquote {\bibinfo {title} {{Heat transport
					in high-Rayleigh-number convection}},}\ }\href@noop {} {\bibfield  {journal}
		{\bibinfo  {journal} {Phys. Rev. A}\ }\textbf {\bibinfo {volume} {42}},\
		\bibinfo {pages} {3650--3653} (\bibinfo {year} {1990})}\BibitemShut {NoStop}%
	\bibitem [{\citenamefont {Cioni}, \citenamefont {Ciliberto},\ and\
		\citenamefont {Sommeria}(1997)}]{Cioni:JFM1997}%
	\BibitemOpen
	\bibfield  {author} {\bibinfo {author} {\bibfnamefont {S.}~\bibnamefont
			{Cioni}}, \bibinfo {author} {\bibfnamefont {S.}~\bibnamefont {Ciliberto}}, \
		and\ \bibinfo {author} {\bibfnamefont {J.}~\bibnamefont {Sommeria}},\
	}\bibfield  {title} {\enquote {\bibinfo {title} {{Strongly turbulent
					Rayleigh{\textendash}B{\'e}nard convection in mercury: comparison with
					results at moderate Prandtl number}},}\ }\href@noop {} {\bibfield  {journal}
		{\bibinfo  {journal} {J. Fluid Mech.}\ }\textbf {\bibinfo {volume} {335}},\
		\bibinfo {pages} {111--140} (\bibinfo {year} {1997})}\BibitemShut {NoStop}%
	\bibitem [{\citenamefont {Scheel}\ and\ \citenamefont
		{Schumacher}(2017)}]{Scheel:PRF2017}%
	\BibitemOpen
	\bibfield  {author} {\bibinfo {author} {\bibfnamefont {J.~D.}\ \bibnamefont
			{Scheel}}\ and\ \bibinfo {author} {\bibfnamefont {J.}~\bibnamefont
			{Schumacher}},\ }\bibfield  {title} {\enquote {\bibinfo {title} {{Predicting
					transition ranges to fully turbulent viscous boundary layers in low Prandtl
					number convection flows}},}\ }\href@noop {} {\bibfield  {journal} {\bibinfo
			{journal} {Phys. Rev. Fluids}\ }\textbf {\bibinfo {volume} {2}},\ \bibinfo
		{pages} {123501} (\bibinfo {year} {2017})}\BibitemShut {NoStop}%
	\bibitem [{\citenamefont {Castaing}\ \emph {et~al.}(1989)\citenamefont
		{Castaing}, \citenamefont {Gunaratne}, \citenamefont {{Kadanoff, L. P.}},
		\citenamefont {Libchaber},\ and\ \citenamefont {Heslot}}]{Castaing:JFM1989}%
	\BibitemOpen
	\bibfield  {author} {\bibinfo {author} {\bibfnamefont {B.}~\bibnamefont
			{Castaing}}, \bibinfo {author} {\bibfnamefont {G.}~\bibnamefont {Gunaratne}},
		\bibinfo {author} {\bibnamefont {{Kadanoff, L. P.}}}, \bibinfo {author}
		{\bibfnamefont {A.}~\bibnamefont {Libchaber}}, \ and\ \bibinfo {author}
		{\bibfnamefont {F.}~\bibnamefont {Heslot}},\ }\bibfield  {title} {\enquote
		{\bibinfo {title} {{Scaling of hard thermal turbulence in Rayleigh-B{\'e}nard
					convection}},}\ }\href@noop {} {\bibfield  {journal} {\bibinfo  {journal} {J.
				Fluid Mech.}\ }\textbf {\bibinfo {volume} {204}},\ \bibinfo {pages} {1--30}
		(\bibinfo {year} {1989})}\BibitemShut {NoStop}%
	\bibitem [{\citenamefont {Chavanne}\ \emph {et~al.}(1997)\citenamefont
		{Chavanne}, \citenamefont {Chill{\`a}}, \citenamefont {Castaing},
		\citenamefont {Hebral}, \citenamefont {Chabaud},\ and\ \citenamefont
		{Chaussy}}]{Chavanne:PRL1997}%
	\BibitemOpen
	\bibfield  {author} {\bibinfo {author} {\bibfnamefont {X.}~\bibnamefont
			{Chavanne}}, \bibinfo {author} {\bibfnamefont {F.}~\bibnamefont
			{Chill{\`a}}}, \bibinfo {author} {\bibfnamefont {B.}~\bibnamefont
			{Castaing}}, \bibinfo {author} {\bibfnamefont {B.}~\bibnamefont {Hebral}},
		\bibinfo {author} {\bibfnamefont {B.}~\bibnamefont {Chabaud}}, \ and\
		\bibinfo {author} {\bibfnamefont {J.}~\bibnamefont {Chaussy}},\ }\bibfield
	{title} {\enquote {\bibinfo {title} {{Observation of the ultimate regime in
					Rayleigh-B{\'e}nard convection}},}\ }\href@noop {} {\bibfield  {journal}
		{\bibinfo  {journal} {Phys. Rev. Lett.}\ }\textbf {\bibinfo {volume} {79}},\
		\bibinfo {pages} {3648--3651} (\bibinfo {year} {1997})}\BibitemShut {NoStop}%
	\bibitem [{\citenamefont {Horn}, \citenamefont {Shishkina},\ and\ \citenamefont
		{Wagner}(2013)}]{Horn:JFM2013}%
	\BibitemOpen
	\bibfield  {author} {\bibinfo {author} {\bibfnamefont {S.}~\bibnamefont
			{Horn}}, \bibinfo {author} {\bibfnamefont {O.}~\bibnamefont {Shishkina}}, \
		and\ \bibinfo {author} {\bibfnamefont {C.}~\bibnamefont {Wagner}},\
	}\bibfield  {title} {\enquote {\bibinfo {title} {{On
					non-Oberbeck–Boussinesq effects in three-dimensional Rayleigh–B{\'e}nard
					convection in glycerol}},}\ }\href@noop {} {\bibfield  {journal} {\bibinfo
			{journal} {J. Fluid Mech.}\ }\textbf {\bibinfo {volume} {724}},\ \bibinfo
		{pages} {175--202} (\bibinfo {year} {2013})}\BibitemShut {NoStop}%
	\bibitem [{\citenamefont {Wagner}\ and\ \citenamefont
		{Shishkina}(2013)}]{Wagner:PF2013}%
	\BibitemOpen
	\bibfield  {author} {\bibinfo {author} {\bibfnamefont {S.}~\bibnamefont
			{Wagner}}\ and\ \bibinfo {author} {\bibfnamefont {O.}~\bibnamefont
			{Shishkina}},\ }\bibfield  {title} {\enquote {\bibinfo {title} {{Aspect-ratio
					dependency of Rayleigh-B{\'e}nard convection in box-shaped containers}},}\
	}\href@noop {} {\bibfield  {journal} {\bibinfo  {journal} {Phys. Fluids}\
		}\textbf {\bibinfo {volume} {25}},\ \bibinfo {pages} {085110} (\bibinfo
		{year} {2013})}\BibitemShut {NoStop}%
	\bibitem [{\citenamefont {Kaczorowski}\ and\ \citenamefont
		{Xia}(2013)}]{Kaczorowski:JFM2013}%
	\BibitemOpen
	\bibfield  {author} {\bibinfo {author} {\bibfnamefont {M.}~\bibnamefont
			{Kaczorowski}}\ and\ \bibinfo {author} {\bibfnamefont {K.-Q.}\ \bibnamefont
			{Xia}},\ }\bibfield  {title} {\enquote {\bibinfo {title} {{Turbulent flow in
					the bulk of Rayleigh{\textendash}B{\'e}nard convection: small-scale
					properties in a cubic cell}},}\ }\href@noop {} {\bibfield  {journal}
		{\bibinfo  {journal} {J. Fluid Mech.}\ }\textbf {\bibinfo {volume} {722}},\
		\bibinfo {pages} {596--617} (\bibinfo {year} {2013})}\BibitemShut {NoStop}%
	\bibitem [{\citenamefont {Niemela}\ and\ \citenamefont
		{Sreenivasan}(2003)}]{Niemela:JFM2003}%
	\BibitemOpen
	\bibfield  {author} {\bibinfo {author} {\bibfnamefont {J.~J.}\ \bibnamefont
			{Niemela}}\ and\ \bibinfo {author} {\bibfnamefont {K.~R.}\ \bibnamefont
			{Sreenivasan}},\ }\bibfield  {title} {\enquote {\bibinfo {title} {{Confined
					turbulent convection}},}\ }\href@noop {} {\bibfield  {journal} {\bibinfo
			{journal} {J. Fluid Mech.}\ }\textbf {\bibinfo {volume} {481}},\ \bibinfo
		{pages} {355--384} (\bibinfo {year} {2003})}\BibitemShut {NoStop}%
	\bibitem [{\citenamefont {Funfschilling}\ \emph {et~al.}(2005)\citenamefont
		{Funfschilling}, \citenamefont {Brown}, \citenamefont {Nikolaenko},\ and\
		\citenamefont {Ahlers}}]{Funfschilling:JFM2005}%
	\BibitemOpen
	\bibfield  {author} {\bibinfo {author} {\bibfnamefont {D.}~\bibnamefont
			{Funfschilling}}, \bibinfo {author} {\bibfnamefont {E.}~\bibnamefont
			{Brown}}, \bibinfo {author} {\bibfnamefont {A.}~\bibnamefont {Nikolaenko}}, \
		and\ \bibinfo {author} {\bibfnamefont {G.}~\bibnamefont {Ahlers}},\
	}\bibfield  {title} {\enquote {\bibinfo {title} {{Heat transport in turbulent
					Rayleigh-B{\'e}nard convection in cylindrical samples with aspect ratio one
					and larger}},}\ }\href@noop {} {\bibfield  {journal} {\bibinfo  {journal} {J.
				Fluid Mech.}\ }\textbf {\bibinfo {volume} {536}},\ \bibinfo {pages}
		{145--154} (\bibinfo {year} {2005})}\BibitemShut {NoStop}%
	\bibitem [{\citenamefont {Stevens}, \citenamefont {Verzicco},\ and\
		\citenamefont {Lohse}(2010)}]{Stevens:JFM2010}%
	\BibitemOpen
	\bibfield  {author} {\bibinfo {author} {\bibfnamefont {R.~J. A.~M.}\
			\bibnamefont {Stevens}}, \bibinfo {author} {\bibfnamefont {R.}~\bibnamefont
			{Verzicco}}, \ and\ \bibinfo {author} {\bibfnamefont {D.}~\bibnamefont
			{Lohse}},\ }\bibfield  {title} {\enquote {\bibinfo {title} {{Radial boundary
					layer structure and Nusselt number in Rayleigh{\textendash}B{\'e}nard
					convection}},}\ }\href@noop {} {\bibfield  {journal} {\bibinfo  {journal} {J.
				Fluid Mech.}\ }\textbf {\bibinfo {volume} {643}},\ \bibinfo {pages}
		{495--507} (\bibinfo {year} {2010})}\BibitemShut {NoStop}%
	\bibitem [{\citenamefont {Dong}\ \emph {et~al.}(2020)\citenamefont {Dong},
		\citenamefont {Wang}, \citenamefont {Dong}, \citenamefont {Huang},
		\citenamefont {Jiang}, \citenamefont {Liu}, \citenamefont {Lu}, \citenamefont
		{Qiu}, \citenamefont {Tang},\ and\ \citenamefont {Zhou}}]{Dong:PF2020}%
	\BibitemOpen
	\bibfield  {author} {\bibinfo {author} {\bibfnamefont {D.-L.}\ \bibnamefont
			{Dong}}, \bibinfo {author} {\bibfnamefont {B.-F.}\ \bibnamefont {Wang}},
		\bibinfo {author} {\bibfnamefont {Y.-H.}\ \bibnamefont {Dong}}, \bibinfo
		{author} {\bibfnamefont {Y.-X.}\ \bibnamefont {Huang}}, \bibinfo {author}
		{\bibfnamefont {N.}~\bibnamefont {Jiang}}, \bibinfo {author} {\bibfnamefont
			{Y.-L.}\ \bibnamefont {Liu}}, \bibinfo {author} {\bibfnamefont {Z.-M.}\
			\bibnamefont {Lu}}, \bibinfo {author} {\bibfnamefont {X.}~\bibnamefont
			{Qiu}}, \bibinfo {author} {\bibfnamefont {Z.-Q.}\ \bibnamefont {Tang}}, \
		and\ \bibinfo {author} {\bibfnamefont {Q.}~\bibnamefont {Zhou}},\ }\bibfield
	{title} {\enquote {\bibinfo {title} {Influence of spatial arrangements of
				roughness elements on turbulent rayleigh-b{\'e}nard convection},}\
	}\href@noop {} {\bibfield  {journal} {\bibinfo  {journal} {Phys. Fluids}\
		}\textbf {\bibinfo {volume} {32}},\ \bibinfo {pages} {045114} (\bibinfo
		{year} {2020})}\BibitemShut {NoStop}%
	\bibitem [{\citenamefont {Madanan}\ and\ \citenamefont
		{Goldstein}(2020)}]{Madanan:PF2020}%
	\BibitemOpen
	\bibfield  {author} {\bibinfo {author} {\bibfnamefont {U.}~\bibnamefont
			{Madanan}}\ and\ \bibinfo {author} {\bibfnamefont {R.~J.}\ \bibnamefont
			{Goldstein}},\ }\bibfield  {title} {\enquote {\bibinfo {title}
			{{High-Rayleigh-number thermal convection of compressed gases in inclined
					rectangular enclosures}},}\ }\href@noop {} {\bibfield  {journal} {\bibinfo
			{journal} {Phys. Fluids}\ }\textbf {\bibinfo {volume} {32}},\ \bibinfo
		{pages} {017103} (\bibinfo {year} {2020})}\BibitemShut {NoStop}%
	\bibitem [{\citenamefont {Vial}\ and\ \citenamefont
		{Hern{\'a}ndez}(2017)}]{Vial:PF2017}%
	\BibitemOpen
	\bibfield  {author} {\bibinfo {author} {\bibfnamefont {M.}~\bibnamefont
			{Vial}}\ and\ \bibinfo {author} {\bibfnamefont {R.~H.}\ \bibnamefont
			{Hern{\'a}ndez}},\ }\bibfield  {title} {\enquote {\bibinfo {title} {{Feedback
					control and heat transfer measurements in a Rayleigh-B{\'e}nard convection
					cell}},}\ }\href@noop {} {\bibfield  {journal} {\bibinfo  {journal} {Phys.
				Fluids}\ }\textbf {\bibinfo {volume} {29}},\ \bibinfo {pages} {074103}
		(\bibinfo {year} {2017})}\BibitemShut {NoStop}%
	\bibitem [{\citenamefont {Xia}, \citenamefont {Lam},\ and\ \citenamefont
		{Zhou}(2002)}]{Xia:PRL2002}%
	\BibitemOpen
	\bibfield  {author} {\bibinfo {author} {\bibfnamefont {K.-Q.}\ \bibnamefont
			{Xia}}, \bibinfo {author} {\bibfnamefont {S.}~\bibnamefont {Lam}}, \ and\
		\bibinfo {author} {\bibfnamefont {S.-Q.}\ \bibnamefont {Zhou}},\ }\bibfield
	{title} {\enquote {\bibinfo {title} {{Heat-flux measurement in
					high-Prandtl-number turbulent Rayleigh-B{\'e}nard convection}},}\ }\href@noop
	{} {\bibfield  {journal} {\bibinfo  {journal} {Phys. Rev. Lett.}\ }\textbf
		{\bibinfo {volume} {88}},\ \bibinfo {pages} {064501} (\bibinfo {year}
		{2002})}\BibitemShut {NoStop}%
	\bibitem [{\citenamefont {Verzicco}\ and\ \citenamefont
		{Camussi}(1999)}]{Verzicco:JFM1999}%
	\BibitemOpen
	\bibfield  {author} {\bibinfo {author} {\bibfnamefont {R.}~\bibnamefont
			{Verzicco}}\ and\ \bibinfo {author} {\bibfnamefont {R.}~\bibnamefont
			{Camussi}},\ }\bibfield  {title} {\enquote {\bibinfo {title} {{Prandtl number
					effects in convective turbulence}},}\ }\href@noop {} {\bibfield  {journal}
		{\bibinfo  {journal} {J. Fluid Mech.}\ }\textbf {\bibinfo {volume} {383}},\
		\bibinfo {pages} {55--73} (\bibinfo {year} {1999})}\BibitemShut {NoStop}%
	\bibitem [{\citenamefont {Niemela}\ \emph {et~al.}(2001)\citenamefont
		{Niemela}, \citenamefont {Skrbek}, \citenamefont {Sreenivasan},\ and\
		\citenamefont {Donnelly}}]{Niemela:JFM2001}%
	\BibitemOpen
	\bibfield  {author} {\bibinfo {author} {\bibfnamefont {J.~J.}\ \bibnamefont
			{Niemela}}, \bibinfo {author} {\bibfnamefont {L.}~\bibnamefont {Skrbek}},
		\bibinfo {author} {\bibfnamefont {K.~R.}\ \bibnamefont {Sreenivasan}}, \ and\
		\bibinfo {author} {\bibfnamefont {R.~J.}\ \bibnamefont {Donnelly}},\
	}\bibfield  {title} {\enquote {\bibinfo {title} {{The wind in confined
					thermal convection}},}\ }\href@noop {} {\bibfield  {journal} {\bibinfo
			{journal} {J. Fluid Mech.}\ }\textbf {\bibinfo {volume} {449}},\ \bibinfo
		{pages} {169--178} (\bibinfo {year} {2001})}\BibitemShut {NoStop}%
	\bibitem [{\citenamefont {Lam}\ \emph {et~al.}(2002)\citenamefont {Lam},
		\citenamefont {Shang}, \citenamefont {Zhou},\ and\ \citenamefont
		{Xia}}]{Lam:PRE2002}%
	\BibitemOpen
	\bibfield  {author} {\bibinfo {author} {\bibfnamefont {S.}~\bibnamefont
			{Lam}}, \bibinfo {author} {\bibfnamefont {X.-D.}\ \bibnamefont {Shang}},
		\bibinfo {author} {\bibfnamefont {S.-Q.}\ \bibnamefont {Zhou}}, \ and\
		\bibinfo {author} {\bibfnamefont {K.-Q.}\ \bibnamefont {Xia}},\ }\bibfield
	{title} {\enquote {\bibinfo {title} {{Prandtl number dependence of the
					viscous boundary layer and the Reynolds numbers in Rayleigh-B{\'e}nard
					convection}},}\ }\href@noop {} {\bibfield  {journal} {\bibinfo  {journal}
			{Phys. Rev. E}\ }\textbf {\bibinfo {volume} {65}},\ \bibinfo {pages} {066306}
		(\bibinfo {year} {2002})}\BibitemShut {NoStop}%
	\bibitem [{\citenamefont {Emran}\ and\ \citenamefont
		{Schumacher}(2008)}]{Emran:JFM2008}%
	\BibitemOpen
	\bibfield  {author} {\bibinfo {author} {\bibfnamefont {M.~S.}\ \bibnamefont
			{Emran}}\ and\ \bibinfo {author} {\bibfnamefont {J.}~\bibnamefont
			{Schumacher}},\ }\bibfield  {title} {\enquote {\bibinfo {title} {{Fine-scale
					statistics of temperature and its derivatives in convective turbulence}},}\
	}\href@noop {} {\bibfield  {journal} {\bibinfo  {journal} {J. Fluid Mech.}\
		}\textbf {\bibinfo {volume} {611}},\ \bibinfo {pages} {13--34} (\bibinfo
		{year} {2008})}\BibitemShut {NoStop}%
	\bibitem [{\citenamefont {Silano}, \citenamefont {Sreenivasan},\ and\
		\citenamefont {Verzicco}(2010)}]{Silano:JFM2010}%
	\BibitemOpen
	\bibfield  {author} {\bibinfo {author} {\bibfnamefont {G.}~\bibnamefont
			{Silano}}, \bibinfo {author} {\bibfnamefont {K.~R.}\ \bibnamefont
			{Sreenivasan}}, \ and\ \bibinfo {author} {\bibfnamefont {R.}~\bibnamefont
			{Verzicco}},\ }\bibfield  {title} {\enquote {\bibinfo {title} {{Numerical
					simulations of Rayleigh{\textendash}B{\'e}nard convection for Prandtl numbers
					between 10$^{-1}$ and 10$^{4}$ and Rayleigh numbers between 10$^{5}$ and
					10$^{9}$}},}\ }\href@noop {} {\bibfield  {journal} {\bibinfo  {journal} {J.
				Fluid Mech.}\ }\textbf {\bibinfo {volume} {662}},\ \bibinfo {pages}
		{409--446} (\bibinfo {year} {2010})}\BibitemShut {NoStop}%
	\bibitem [{\citenamefont {Verma}\ \emph {et~al.}(2012)\citenamefont {Verma},
		\citenamefont {Mishra}, \citenamefont {Pandey},\ and\ \citenamefont
		{Paul}}]{Verma:PRE2012}%
	\BibitemOpen
	\bibfield  {author} {\bibinfo {author} {\bibfnamefont {M.~K.}\ \bibnamefont
			{Verma}}, \bibinfo {author} {\bibfnamefont {P.~K.}\ \bibnamefont {Mishra}},
		\bibinfo {author} {\bibfnamefont {A.}~\bibnamefont {Pandey}}, \ and\ \bibinfo
		{author} {\bibfnamefont {S.}~\bibnamefont {Paul}},\ }\bibfield  {title}
	{\enquote {\bibinfo {title} {{Scalings of field correlations and heat
					transport in turbulent convection}},}\ }\href@noop {} {\bibfield  {journal}
		{\bibinfo  {journal} {Phys. Rev. E}\ }\textbf {\bibinfo {volume} {85}},\
		\bibinfo {pages} {016310} (\bibinfo {year} {2012})}\BibitemShut {NoStop}%
	\bibitem [{\citenamefont {Pandey}\ and\ \citenamefont
		{Verma}(2016)}]{Pandey:PF2016}%
	\BibitemOpen
	\bibfield  {author} {\bibinfo {author} {\bibfnamefont {A.}~\bibnamefont
			{Pandey}}\ and\ \bibinfo {author} {\bibfnamefont {M.~K.}\ \bibnamefont
			{Verma}},\ }\bibfield  {title} {\enquote {\bibinfo {title} {{Scaling of
					large-scale quantities in Rayleigh-B{\'e}nard convection}},}\ }\href@noop {}
	{\bibfield  {journal} {\bibinfo  {journal} {Phys. Fluids}\ }\textbf {\bibinfo
			{volume} {28}},\ \bibinfo {pages} {095105} (\bibinfo {year}
		{2016})}\BibitemShut {NoStop}%
	\bibitem [{\citenamefont {Pandey}\ \emph {et~al.}(2016)\citenamefont {Pandey},
		\citenamefont {Kumar}, \citenamefont {Chatterjee},\ and\ \citenamefont
		{Verma}}]{Pandey:PRE2016}%
	\BibitemOpen
	\bibfield  {author} {\bibinfo {author} {\bibfnamefont {A.}~\bibnamefont
			{Pandey}}, \bibinfo {author} {\bibfnamefont {A.}~\bibnamefont {Kumar}},
		\bibinfo {author} {\bibfnamefont {A.~G.}\ \bibnamefont {Chatterjee}}, \ and\
		\bibinfo {author} {\bibfnamefont {M.~K.}\ \bibnamefont {Verma}},\ }\bibfield
	{title} {\enquote {\bibinfo {title} {{Dynamics of large-scale quantities in
					Rayleigh-B{\'e}nard convection}},}\ }\href@noop {} {\bibfield  {journal}
		{\bibinfo  {journal} {Phys. Rev. E}\ }\textbf {\bibinfo {volume} {94}},\
		\bibinfo {pages} {053106} (\bibinfo {year} {2016})}\BibitemShut {NoStop}%
	\bibitem [{\citenamefont {Brown}, \citenamefont {Funfschilling},\ and\
		\citenamefont {Ahlers}(2007)}]{Brown:JSM2007}%
	\BibitemOpen
	\bibfield  {author} {\bibinfo {author} {\bibfnamefont {E.}~\bibnamefont
			{Brown}}, \bibinfo {author} {\bibfnamefont {D.}~\bibnamefont
			{Funfschilling}}, \ and\ \bibinfo {author} {\bibfnamefont {G.}~\bibnamefont
			{Ahlers}},\ }\bibfield  {title} {\enquote {\bibinfo {title} {{Anomalous
					Reynolds-number scaling in turbulent Rayleigh{\textendash}B{\'e}nard
					convection}},}\ }\href@noop {} {\bibfield  {journal} {\bibinfo  {journal} {J.
				Stat. Mech. Theor. Exp.}\ }\textbf {\bibinfo {volume} {2007}},\ \bibinfo
		{pages} {P10005} (\bibinfo {year} {2007})}\BibitemShut {NoStop}%
	\bibitem [{\citenamefont {Kraichnan}(1962)}]{Kraichnan:PF1962Convection}%
	\BibitemOpen
	\bibfield  {author} {\bibinfo {author} {\bibfnamefont {R.~H.}\ \bibnamefont
			{Kraichnan}},\ }\bibfield  {title} {\enquote {\bibinfo {title} {{Turbulent
					thermal convection at arbitrary prandtl number}},}\ }\href@noop {} {\bibfield
		{journal} {\bibinfo  {journal} {Phys. Fluids}\ }\textbf {\bibinfo {volume}
			{5}},\ \bibinfo {pages} {1374--1389} (\bibinfo {year} {1962})}\BibitemShut
	{NoStop}%
	\bibitem [{\citenamefont {Lohse}\ and\ \citenamefont
		{Toschi}(2003)}]{Lohse:PRL2003}%
	\BibitemOpen
	\bibfield  {author} {\bibinfo {author} {\bibfnamefont {D.}~\bibnamefont
			{Lohse}}\ and\ \bibinfo {author} {\bibfnamefont {F.}~\bibnamefont {Toschi}},\
	}\bibfield  {title} {\enquote {\bibinfo {title} {{Ultimate state of thermal
					convection}},}\ }\href@noop {} {\bibfield  {journal} {\bibinfo  {journal}
			{Phys. Rev. Lett.}\ }\textbf {\bibinfo {volume} {90}},\ \bibinfo {pages}
		{034502} (\bibinfo {year} {2003})}\BibitemShut {NoStop}%
	\bibitem [{\citenamefont {Pawar}\ and\ \citenamefont
		{Arakeri}(2016)}]{Pawar:PRF2016}%
	\BibitemOpen
	\bibfield  {author} {\bibinfo {author} {\bibfnamefont {S.~S.}\ \bibnamefont
			{Pawar}}\ and\ \bibinfo {author} {\bibfnamefont {J.~H.}\ \bibnamefont
			{Arakeri}},\ }\bibfield  {title} {\enquote {\bibinfo {title} {{Two regimes of
					flux scaling in axially homogeneous turbulent convection in vertical
					tube}},}\ }\href@noop {} {\bibfield  {journal} {\bibinfo  {journal} {Phys.
				Rev. Fluids}\ }\textbf {\bibinfo {volume} {1}},\ \bibinfo {pages} {042401(R)}
		(\bibinfo {year} {2016})}\BibitemShut {NoStop}%
	\bibitem [{\citenamefont {Schmidt}\ \emph {et~al.}(2012)\citenamefont
		{Schmidt}, \citenamefont {Calzavarini}, \citenamefont {Lohse},\ and\
		\citenamefont {Toschi}}]{Schmidt:JFM2012}%
	\BibitemOpen
	\bibfield  {author} {\bibinfo {author} {\bibfnamefont {L.~E.}\ \bibnamefont
			{Schmidt}}, \bibinfo {author} {\bibfnamefont {E.}~\bibnamefont
			{Calzavarini}}, \bibinfo {author} {\bibfnamefont {D.}~\bibnamefont {Lohse}},
		\ and\ \bibinfo {author} {\bibfnamefont {F.}~\bibnamefont {Toschi}},\
	}\bibfield  {title} {\enquote {\bibinfo {title} {{Axially homogeneous
					Rayleigh-B{\'e}nard convection in a cylindrical cell}},}\ }\href@noop {}
	{\bibfield  {journal} {\bibinfo  {journal} {J. Fluid Mech.}\ }\textbf
		{\bibinfo {volume} {691}},\ \bibinfo {pages} {52--68} (\bibinfo {year}
		{2012})}\BibitemShut {NoStop}%
	\bibitem [{\citenamefont {Roche}\ \emph {et~al.}(2001)\citenamefont {Roche},
		\citenamefont {Castaing}, \citenamefont {Chabaud},\ and\ \citenamefont
		{Hebral}}]{Roche:PRE2001}%
	\BibitemOpen
	\bibfield  {author} {\bibinfo {author} {\bibfnamefont {P.-E.}\ \bibnamefont
			{Roche}}, \bibinfo {author} {\bibfnamefont {B.}~\bibnamefont {Castaing}},
		\bibinfo {author} {\bibfnamefont {B.}~\bibnamefont {Chabaud}}, \ and\
		\bibinfo {author} {\bibfnamefont {B.}~\bibnamefont {Hebral}},\ }\bibfield
	{title} {\enquote {\bibinfo {title} {{Observation of the 1/2 power law in
					Rayleigh-B{\'e}nard convection}},}\ }\href@noop {} {\bibfield  {journal}
		{\bibinfo  {journal} {Phys. Rev. E}\ }\textbf {\bibinfo {volume} {63}},\
		\bibinfo {pages} {045303(R)} (\bibinfo {year} {2001})}\BibitemShut {NoStop}%
	\bibitem [{\citenamefont {Ahlers}\ \emph {et~al.}(2012)\citenamefont {Ahlers},
		\citenamefont {He}, \citenamefont {Funfschilling},\ and\ \citenamefont
		{Bodenschatz}}]{Ahlers:NJP2012}%
	\BibitemOpen
	\bibfield  {author} {\bibinfo {author} {\bibfnamefont {G.}~\bibnamefont
			{Ahlers}}, \bibinfo {author} {\bibfnamefont {X.}~\bibnamefont {He}}, \bibinfo
		{author} {\bibfnamefont {D.}~\bibnamefont {Funfschilling}}, \ and\ \bibinfo
		{author} {\bibfnamefont {E.}~\bibnamefont {Bodenschatz}},\ }\bibfield
	{title} {\enquote {\bibinfo {title} {{Heat transport by turbulent
					Rayleigh-B{\'e}nard convection for $Pr \approx 0.8$ and $3 \times 10^{12}
					\lesssim Ra \lesssim 10^{15}$: aspect ratio T = 0.50}},}\ }\href@noop {}
	{\bibfield  {journal} {\bibinfo  {journal} {New J. Phys.}\ }\textbf {\bibinfo
			{volume} {12}},\ \bibinfo {pages} {103012} (\bibinfo {year}
		{2012})}\BibitemShut {NoStop}%
	\bibitem [{\citenamefont {He}\ \emph {et~al.}(2012)\citenamefont {He},
		\citenamefont {Funfschilling}, \citenamefont {Nobach}, \citenamefont
		{Bodenschatz},\ and\ \citenamefont {Ahlers}}]{He:PRL2012}%
	\BibitemOpen
	\bibfield  {author} {\bibinfo {author} {\bibfnamefont {X.}~\bibnamefont
			{He}}, \bibinfo {author} {\bibfnamefont {D.}~\bibnamefont {Funfschilling}},
		\bibinfo {author} {\bibfnamefont {H.}~\bibnamefont {Nobach}}, \bibinfo
		{author} {\bibfnamefont {E.}~\bibnamefont {Bodenschatz}}, \ and\ \bibinfo
		{author} {\bibfnamefont {G.}~\bibnamefont {Ahlers}},\ }\bibfield  {title}
	{\enquote {\bibinfo {title} {{Transition to the Ultimate State of Turbulent
					Rayleigh-B{\'e}nard Convection}},}\ }\href@noop {} {\bibfield  {journal}
		{\bibinfo  {journal} {Phys. Rev. Lett.}\ }\textbf {\bibinfo {volume} {108}},\
		\bibinfo {pages} {024502} (\bibinfo {year} {2012})}\BibitemShut {NoStop}%
	\bibitem [{\citenamefont {Niemela}\ \emph {et~al.}(2000)\citenamefont
		{Niemela}, \citenamefont {Skrbek}, \citenamefont {Sreenivasan},\ and\
		\citenamefont {Donnelly}}]{Niemela:Nature2000}%
	\BibitemOpen
	\bibfield  {author} {\bibinfo {author} {\bibfnamefont {J.~J.}\ \bibnamefont
			{Niemela}}, \bibinfo {author} {\bibfnamefont {L.}~\bibnamefont {Skrbek}},
		\bibinfo {author} {\bibfnamefont {K.~R.}\ \bibnamefont {Sreenivasan}}, \ and\
		\bibinfo {author} {\bibfnamefont {R.~J.}\ \bibnamefont {Donnelly}},\
	}\bibfield  {title} {\enquote {\bibinfo {title} {{Turbulent convection at
					very high Rayleigh numbers}},}\ }\href@noop {} {\bibfield  {journal}
		{\bibinfo  {journal} {Nature}\ }\textbf {\bibinfo {volume} {404}},\ \bibinfo
		{pages} {837--840} (\bibinfo {year} {2000})}\BibitemShut {NoStop}%
	\bibitem [{\citenamefont {Urban}\ \emph {et~al.}(2012)\citenamefont {Urban},
		\citenamefont {Hanzelka}, \citenamefont {Kralik}, \citenamefont
		{Musilov{\'a}}, \citenamefont {Srnka},\ and\ \citenamefont
		{Skrbek}}]{Urban:PRL2012}%
	\BibitemOpen
	\bibfield  {author} {\bibinfo {author} {\bibfnamefont {P.}~\bibnamefont
			{Urban}}, \bibinfo {author} {\bibfnamefont {P.}~\bibnamefont {Hanzelka}},
		\bibinfo {author} {\bibfnamefont {T.}~\bibnamefont {Kralik}}, \bibinfo
		{author} {\bibfnamefont {V.}~\bibnamefont {Musilov{\'a}}}, \bibinfo {author}
		{\bibfnamefont {A.}~\bibnamefont {Srnka}}, \ and\ \bibinfo {author}
		{\bibfnamefont {L.}~\bibnamefont {Skrbek}},\ }\bibfield  {title} {\enquote
		{\bibinfo {title} {{Effect of Boundary Layers Asymmetry on Heat Transfer
					Efficiency in Turbulent Rayleigh-Bernard Convection at Very High Rayleigh
					Numbers}},}\ }\href@noop {} {\bibfield  {journal} {\bibinfo  {journal} {Phys.
				Rev. Lett.}\ }\textbf {\bibinfo {volume} {109}},\ \bibinfo {pages} {154301}
		(\bibinfo {year} {2012})}\BibitemShut {NoStop}%
	\bibitem [{\citenamefont {Grossmann}\ and\ \citenamefont
		{Lohse}(2000)}]{Grossmann:JFM2000}%
	\BibitemOpen
	\bibfield  {author} {\bibinfo {author} {\bibfnamefont {S.}~\bibnamefont
			{Grossmann}}\ and\ \bibinfo {author} {\bibfnamefont {D.}~\bibnamefont
			{Lohse}},\ }\bibfield  {title} {\enquote {\bibinfo {title} {{Scaling in
					thermal convection: a unifying theory}},}\ }\href@noop {} {\bibfield
		{journal} {\bibinfo  {journal} {J. Fluid Mech.}\ }\textbf {\bibinfo {volume}
			{407}},\ \bibinfo {pages} {27--56} (\bibinfo {year} {2000})}\BibitemShut
	{NoStop}%
	\bibitem [{\citenamefont {Grossmann}\ and\ \citenamefont
		{Lohse}(2001)}]{Grossmann:PRL2001}%
	\BibitemOpen
	\bibfield  {author} {\bibinfo {author} {\bibfnamefont {S.}~\bibnamefont
			{Grossmann}}\ and\ \bibinfo {author} {\bibfnamefont {D.}~\bibnamefont
			{Lohse}},\ }\bibfield  {title} {\enquote {\bibinfo {title} {{Thermal
					convection for large Prandtl numbers}},}\ }\href@noop {} {\bibfield
		{journal} {\bibinfo  {journal} {Phys. Rev. Lett.}\ }\textbf {\bibinfo
			{volume} {86}},\ \bibinfo {pages} {3316--3319} (\bibinfo {year}
		{2001})}\BibitemShut {NoStop}%
	\bibitem [{\citenamefont {Bhattacharya}, \citenamefont {Verma},\ and\
		\citenamefont {Samtaney}(2021{\natexlab{a}})}]{Bhattacharya:PF2021}%
	\BibitemOpen
	\bibfield  {author} {\bibinfo {author} {\bibfnamefont {S.}~\bibnamefont
			{Bhattacharya}}, \bibinfo {author} {\bibfnamefont {M.~K.}\ \bibnamefont
			{Verma}}, \ and\ \bibinfo {author} {\bibfnamefont {R.}~\bibnamefont
			{Samtaney}},\ }\bibfield  {title} {\enquote {\bibinfo {title} {{Revisiting
					Reynolds and Nusselt numbers in turbulent thermal convection}},}\ }\href@noop
	{} {\bibfield  {journal} {\bibinfo  {journal} {Phys. Fluids}\ }\textbf
		{\bibinfo {volume} {33}},\ \bibinfo {pages} {015113} (\bibinfo {year}
		{2021}{\natexlab{a}})}\BibitemShut {NoStop}%
	\bibitem [{\citenamefont {Bhattacharya}\ \emph {et~al.}(2018)\citenamefont
		{Bhattacharya}, \citenamefont {Pandey}, \citenamefont {Kumar},\ and\
		\citenamefont {Verma}}]{Bhattacharya:PF2018}%
	\BibitemOpen
	\bibfield  {author} {\bibinfo {author} {\bibfnamefont {S.}~\bibnamefont
			{Bhattacharya}}, \bibinfo {author} {\bibfnamefont {A.}~\bibnamefont
			{Pandey}}, \bibinfo {author} {\bibfnamefont {A.}~\bibnamefont {Kumar}}, \
		and\ \bibinfo {author} {\bibfnamefont {M.~K.}\ \bibnamefont {Verma}},\
	}\bibfield  {title} {\enquote {\bibinfo {title} {{Complexity of viscous
					dissipation in turbulent thermal convection}},}\ }\href@noop {} {\bibfield
		{journal} {\bibinfo  {journal} {Phys. Fluids}\ }\textbf {\bibinfo {volume}
			{30}},\ \bibinfo {pages} {031702} (\bibinfo {year} {2018})}\BibitemShut
	{NoStop}%
	\bibitem [{\citenamefont {Bhattacharya}, \citenamefont {Samtaney},\ and\
		\citenamefont {Verma}(2019)}]{Bhattacharya:PF2019}%
	\BibitemOpen
	\bibfield  {author} {\bibinfo {author} {\bibfnamefont {S.}~\bibnamefont
			{Bhattacharya}}, \bibinfo {author} {\bibfnamefont {R.}~\bibnamefont
			{Samtaney}}, \ and\ \bibinfo {author} {\bibfnamefont {M.~K.}\ \bibnamefont
			{Verma}},\ }\bibfield  {title} {\enquote {\bibinfo {title} {{Scaling and
					spatial intermittency of thermal dissipation in turbulent convection}},}\
	}\href@noop {} {\bibfield  {journal} {\bibinfo  {journal} {Phys. Fluids}\
		}\textbf {\bibinfo {volume} {31}},\ \bibinfo {pages} {075104} (\bibinfo
		{year} {2019})}\BibitemShut {NoStop}%
	\bibitem [{\citenamefont {Parish}\ and\ \citenamefont
		{Duraisamy}(2016)}]{Parish:JCP2016}%
	\BibitemOpen
	\bibfield  {author} {\bibinfo {author} {\bibfnamefont {E.~J.}\ \bibnamefont
			{Parish}}\ and\ \bibinfo {author} {\bibfnamefont {K.}~\bibnamefont
			{Duraisamy}},\ }\bibfield  {title} {\enquote {\bibinfo {title} {{A paradigm
					for data-driven predictive modeling using field inversion and machine
					learning}},}\ }\href@noop {} {\bibfield  {journal} {\bibinfo  {journal} {J.
				Comput. Phys.}\ }\textbf {\bibinfo {volume} {305}},\ \bibinfo {pages}
		{758--774} (\bibinfo {year} {2016})}\BibitemShut {NoStop}%
	\bibitem [{\citenamefont {Fonda}\ \emph {et~al.}(2019)\citenamefont {Fonda},
		\citenamefont {Pandey}, \citenamefont {Schumacher},\ and\ \citenamefont
		{Sreenivasan}}]{Fonda:PNAS2019}%
	\BibitemOpen
	\bibfield  {author} {\bibinfo {author} {\bibfnamefont {E.}~\bibnamefont
			{Fonda}}, \bibinfo {author} {\bibfnamefont {A.}~\bibnamefont {Pandey}},
		\bibinfo {author} {\bibfnamefont {J.}~\bibnamefont {Schumacher}}, \ and\
		\bibinfo {author} {\bibfnamefont {K.~R.}\ \bibnamefont {Sreenivasan}},\
	}\bibfield  {title} {\enquote {\bibinfo {title} {{Deep learning in turbulent
					convection networks}},}\ }\href@noop {} {\bibfield  {journal} {\bibinfo
			{journal} {Proc. Natl. Acad. Sci. U.S.A.}\ }\textbf {\bibinfo {volume}
			{116}},\ \bibinfo {pages} {8667--8672} (\bibinfo {year} {2019})}\BibitemShut
	{NoStop}%
	\bibitem [{\citenamefont {Pandey}\ and\ \citenamefont
		{Schumacher}(2020)}]{Pandey:PRF2020}%
	\BibitemOpen
	\bibfield  {author} {\bibinfo {author} {\bibfnamefont {S.}~\bibnamefont
			{Pandey}}\ and\ \bibinfo {author} {\bibfnamefont {J.}~\bibnamefont
			{Schumacher}},\ }\bibfield  {title} {\enquote {\bibinfo {title} {{Reservoir
					computing model of two-dimensional turbulent convection}},}\ }\href@noop {}
	{\bibfield  {journal} {\bibinfo  {journal} {Phys. Rev. Fluids}\ }\textbf
		{\bibinfo {volume} {5}},\ \bibinfo {pages} {113506} (\bibinfo {year}
		{2020})}\BibitemShut {NoStop}%
	\bibitem [{\citenamefont {Pandey}, \citenamefont {Schumacher},\ and\
		\citenamefont {Sreenivasan}(2020)}]{Pandey:JOT2020}%
	\BibitemOpen
	\bibfield  {author} {\bibinfo {author} {\bibfnamefont {S.}~\bibnamefont
			{Pandey}}, \bibinfo {author} {\bibfnamefont {J.}~\bibnamefont {Schumacher}},
		\ and\ \bibinfo {author} {\bibfnamefont {K.~R.}\ \bibnamefont
			{Sreenivasan}},\ }\bibfield  {title} {\enquote {\bibinfo {title} {{A
					perspective on machine learning in turbulent flows}},}\ }\href@noop {}
	{\bibfield  {journal} {\bibinfo  {journal} {J. of Turbulence}\ }\textbf
		{\bibinfo {volume} {21}},\ \bibinfo {pages} {567--584} (\bibinfo {year}
		{2020})}\BibitemShut {NoStop}%
	\bibitem [{\citenamefont {Brunton}, \citenamefont {Noack},\ and\ \citenamefont
		{Koumoutsakos}(2020)}]{Brunton:ARFM2020}%
	\BibitemOpen
	\bibfield  {author} {\bibinfo {author} {\bibfnamefont {S.~L.}\ \bibnamefont
			{Brunton}}, \bibinfo {author} {\bibfnamefont {B.~R.}\ \bibnamefont {Noack}},
		\ and\ \bibinfo {author} {\bibfnamefont {P.}~\bibnamefont {Koumoutsakos}},\
	}\bibfield  {title} {\enquote {\bibinfo {title} {{Machine Learning for Fluid
					Mechanics}},}\ }\href@noop {} {\bibfield  {journal} {\bibinfo  {journal}
			{Annu. Rev. Fluid Mech.}\ }\textbf {\bibinfo {volume} {52}},\ \bibinfo
		{pages} {477--508} (\bibinfo {year} {2020})}\BibitemShut {NoStop}%
	\bibitem [{\citenamefont {Goodfellow}, \citenamefont {Bengio},\ and\
		\citenamefont {Courville}(2016)}]{Goodfellow:book}%
	\BibitemOpen
	\bibfield  {author} {\bibinfo {author} {\bibfnamefont {I.}~\bibnamefont
			{Goodfellow}}, \bibinfo {author} {\bibfnamefont {Y.}~\bibnamefont {Bengio}},
		\ and\ \bibinfo {author} {\bibfnamefont {A.}~\bibnamefont {Courville}},\
	}\href@noop {} {\emph {\bibinfo {title} {{Deep Learning}}}}\ (\bibinfo
	{publisher} {The MIT Press},\ \bibinfo {address} {Cambridge, Massachusetts},\
	\bibinfo {year} {2016})\BibitemShut {NoStop}%
	\bibitem [{\citenamefont {Hastie}, \citenamefont {Tibshirani},\ and\
		\citenamefont {Friedman}(2009)}]{Hastie:book}%
	\BibitemOpen
	\bibfield  {author} {\bibinfo {author} {\bibfnamefont {T.}~\bibnamefont
			{Hastie}}, \bibinfo {author} {\bibfnamefont {R.}~\bibnamefont {Tibshirani}},
		\ and\ \bibinfo {author} {\bibfnamefont {J.}~\bibnamefont {Friedman}},\
	}\href@noop {} {\emph {\bibinfo {title} {{The Elements of Statistical
					Learning}}}}\ (\bibinfo  {publisher} {Springer},\ \bibinfo {address} {New
		York},\ \bibinfo {year} {2009})\BibitemShut {NoStop}%
	\bibitem [{\citenamefont {Burkov}(2019)}]{Burkov:book}%
	\BibitemOpen
	\bibfield  {author} {\bibinfo {author} {\bibfnamefont {A.}~\bibnamefont
			{Burkov}},\ }\href@noop {} {\emph {\bibinfo {title} {{The Hundred-Page
					Machine Learning Book}}}}\ (\bibinfo  {publisher} {Andriy Burkov},\ \bibinfo
	{address} {Qu{\'e}bec},\ \bibinfo {year} {2019})\BibitemShut {NoStop}%
	\bibitem [{\citenamefont {Stevens}\ \emph {et~al.}(2013)\citenamefont
		{Stevens}, \citenamefont {van~der Poel}, \citenamefont {Grossmann},\ and\
		\citenamefont {Lohse}}]{Stevens:JFM2013}%
	\BibitemOpen
	\bibfield  {author} {\bibinfo {author} {\bibfnamefont {R.~J. A.~M.}\
			\bibnamefont {Stevens}}, \bibinfo {author} {\bibfnamefont {E.~P.}\
			\bibnamefont {van~der Poel}}, \bibinfo {author} {\bibfnamefont
			{S.}~\bibnamefont {Grossmann}}, \ and\ \bibinfo {author} {\bibfnamefont
			{D.}~\bibnamefont {Lohse}},\ }\bibfield  {title} {\enquote {\bibinfo {title}
			{{The unifying theory of scaling in thermal convection: the updated
					prefactors}},}\ }\href@noop {} {\bibfield  {journal} {\bibinfo  {journal} {J.
				Fluid Mech.}\ }\textbf {\bibinfo {volume} {730}},\ \bibinfo {pages}
		{295--308} (\bibinfo {year} {2013})}\BibitemShut {NoStop}%
	\bibitem [{\citenamefont {Lesieur}(2008)}]{Lesieur:book:Turbulence}%
	\BibitemOpen
	\bibfield  {author} {\bibinfo {author} {\bibfnamefont {M.}~\bibnamefont
			{Lesieur}},\ }\href@noop {} {\emph {\bibinfo {title} {{Turbulence in
					Fluids}}}}\ (\bibinfo  {publisher} {Springer-Verlag},\ \bibinfo {address}
	{Dordrecht},\ \bibinfo {year} {2008})\BibitemShut {NoStop}%
	\bibitem [{\citenamefont {Verma}(2019)}]{Verma:book:ET}%
	\BibitemOpen
	\bibfield  {author} {\bibinfo {author} {\bibfnamefont {M.~K.}\ \bibnamefont
			{Verma}},\ }\href@noop {} {\emph {\bibinfo {title} {Energy trasnfers in Fluid
				Flows: Multiscale and Spectral Perspectives}}}\ (\bibinfo  {publisher}
	{Cambridge University Press},\ \bibinfo {address} {Cambridge},\ \bibinfo
	{year} {2019})\BibitemShut {NoStop}%
	\bibitem [{\citenamefont {Bhattacharya}\ \emph {et~al.}(2019)\citenamefont
		{Bhattacharya}, \citenamefont {Sadhukhan}, \citenamefont {Guha},\ and\
		\citenamefont {Verma}}]{Bhattacharya:PF2019b}%
	\BibitemOpen
	\bibfield  {author} {\bibinfo {author} {\bibfnamefont {S.}~\bibnamefont
			{Bhattacharya}}, \bibinfo {author} {\bibfnamefont {S.}~\bibnamefont
			{Sadhukhan}}, \bibinfo {author} {\bibfnamefont {A.}~\bibnamefont {Guha}}, \
		and\ \bibinfo {author} {\bibfnamefont {M.~K.}\ \bibnamefont {Verma}},\
	}\bibfield  {title} {\enquote {\bibinfo {title} {{Similarities between the
					structure functions of thermal convection and hydrodynamic turbulence}},}\
	}\href@noop {} {\bibfield  {journal} {\bibinfo  {journal} {Phys. Fluids}\
		}\textbf {\bibinfo {volume} {31}},\ \bibinfo {pages} {115107} (\bibinfo
		{year} {2019})}\BibitemShut {NoStop}%
	\bibitem [{\citenamefont {Scheel}, \citenamefont {Kim},\ and\ \citenamefont
		{White}(2012)}]{Scheel:JFM2012}%
	\BibitemOpen
	\bibfield  {author} {\bibinfo {author} {\bibfnamefont {J.~D.}\ \bibnamefont
			{Scheel}}, \bibinfo {author} {\bibfnamefont {E.}~\bibnamefont {Kim}}, \ and\
		\bibinfo {author} {\bibfnamefont {K.~R.}\ \bibnamefont {White}},\ }\bibfield
	{title} {\enquote {\bibinfo {title} {{Thermal and viscous boundary layers in
					turbulent Rayleigh{\textendash}B{\'e}nard convection}},}\ }\href@noop {}
	{\bibfield  {journal} {\bibinfo  {journal} {J. Fluid Mech.}\ }\textbf
		{\bibinfo {volume} {711}},\ \bibinfo {pages} {281--305} (\bibinfo {year}
		{2012})}\BibitemShut {NoStop}%
	\bibitem [{\citenamefont {Shi}, \citenamefont {Emran},\ and\ \citenamefont
		{Schumacher}(2012)}]{Shi:JFM2012}%
	\BibitemOpen
	\bibfield  {author} {\bibinfo {author} {\bibfnamefont {N.}~\bibnamefont
			{Shi}}, \bibinfo {author} {\bibfnamefont {M.~S.}\ \bibnamefont {Emran}}, \
		and\ \bibinfo {author} {\bibfnamefont {J.}~\bibnamefont {Schumacher}},\
	}\bibfield  {title} {\enquote {\bibinfo {title} {{Boundary layer structure in
					turbulent Rayleigh{\textendash}B{\'e}nard convection}},}\ }\href@noop {}
	{\bibfield  {journal} {\bibinfo  {journal} {J. Fluid Mech.}\ }\textbf
		{\bibinfo {volume} {706}},\ \bibinfo {pages} {5--33} (\bibinfo {year}
		{2012})}\BibitemShut {NoStop}%
	\bibitem [{\citenamefont {Samuel}\ \emph {et~al.}(2021)\citenamefont {Samuel},
		\citenamefont {Bhattacharya}, \citenamefont {Asad}, \citenamefont
		{Chatterjee}, \citenamefont {Verma}, \citenamefont {Samtaney},\ and\
		\citenamefont {Anwer}}]{Samuel:JOSS2020}%
	\BibitemOpen
	\bibfield  {author} {\bibinfo {author} {\bibfnamefont {R.~J.}\ \bibnamefont
			{Samuel}}, \bibinfo {author} {\bibfnamefont {S.}~\bibnamefont
			{Bhattacharya}}, \bibinfo {author} {\bibfnamefont {A.}~\bibnamefont {Asad}},
		\bibinfo {author} {\bibfnamefont {S.}~\bibnamefont {Chatterjee}}, \bibinfo
		{author} {\bibfnamefont {M.~K.}\ \bibnamefont {Verma}}, \bibinfo {author}
		{\bibfnamefont {R.}~\bibnamefont {Samtaney}}, \ and\ \bibinfo {author}
		{\bibfnamefont {S.~F.}\ \bibnamefont {Anwer}},\ }\bibfield  {title} {\enquote
		{\bibinfo {title} {{SARAS: A general-purpose PDE solver for fluid
					dynamics}},}\ }\href@noop {} {\bibfield  {journal} {\bibinfo  {journal} {J.
				Open Source Softw.}\ }\textbf {\bibinfo {volume} {6}},\ \bibinfo {pages}
		{2095} (\bibinfo {year} {2021})}\BibitemShut {NoStop}%
	\bibitem [{\citenamefont {Verma}\ \emph {et~al.}(2020)\citenamefont {Verma},
		\citenamefont {Samuel}, \citenamefont {Chatterjee}, \citenamefont
		{Bhattacharya},\ and\ \citenamefont {Asad}}]{Verma:SNC2020}%
	\BibitemOpen
	\bibfield  {author} {\bibinfo {author} {\bibfnamefont {M.~K.}\ \bibnamefont
			{Verma}}, \bibinfo {author} {\bibfnamefont {R.~J.}\ \bibnamefont {Samuel}},
		\bibinfo {author} {\bibfnamefont {S.}~\bibnamefont {Chatterjee}}, \bibinfo
		{author} {\bibfnamefont {S.}~\bibnamefont {Bhattacharya}}, \ and\ \bibinfo
		{author} {\bibfnamefont {A.}~\bibnamefont {Asad}},\ }\bibfield  {title}
	{\enquote {\bibinfo {title} {{Challenges in fluid flow simulations using
					exascale computing}},}\ }\href@noop {} {\bibfield  {journal} {\bibinfo
			{journal} {S.N. Comput. Sci.}\ }\textbf {\bibinfo {volume} {1}},\ \bibinfo
		{pages} {178} (\bibinfo {year} {2020})}\BibitemShut {NoStop}%
	\bibitem [{\citenamefont {Bhattacharya}, \citenamefont {Verma},\ and\
		\citenamefont {Samtaney}(2021{\natexlab{b}})}]{Bhattacharya:PRF2021}%
	\BibitemOpen
	\bibfield  {author} {\bibinfo {author} {\bibfnamefont {S.}~\bibnamefont
			{Bhattacharya}}, \bibinfo {author} {\bibfnamefont {M.~K.}\ \bibnamefont
			{Verma}}, \ and\ \bibinfo {author} {\bibfnamefont {R.}~\bibnamefont
			{Samtaney}},\ }\bibfield  {title} {\enquote {\bibinfo {title} {{Prandtl
					number dependence of the small-scale properties in turbulent
					Rayleigh-B{\'e}nard convection}},}\ }\href@noop {} {\bibfield  {journal}
		{\bibinfo  {journal} {Phys. Rev. Fluids}\ }\textbf {\bibinfo {volume} {6}},\
		\bibinfo {pages} {063501} (\bibinfo {year} {2021}{\natexlab{b}})}\BibitemShut
	{NoStop}%
	\bibitem [{\citenamefont {Jekel}\ and\ \citenamefont
		{Venter}(2019)}]{Jekel:PWLF2019}%
	\BibitemOpen
	\bibfield  {author} {\bibinfo {author} {\bibfnamefont {C.~F.}\ \bibnamefont
			{Jekel}}\ and\ \bibinfo {author} {\bibfnamefont {G.}~\bibnamefont {Venter}},\
	}\bibfield  {title} {\enquote {\bibinfo {title} {Pwlf: a python library for
				fitting 1d continuous piecewise linear functions},}\ }\href@noop {}
	{\bibfield  {journal} {\bibinfo  {journal} {URL: https://github.
				com/cjekel/piecewise\_linear\_fit\_py}\ } (\bibinfo {year}
		{2019})}\BibitemShut {NoStop}%
	\bibitem [{\citenamefont {Frank}\ \emph {et~al.}(2009)\citenamefont {Frank},
		\citenamefont {Hall}, \citenamefont {Holmes}, \citenamefont {Kirkby},
		\citenamefont {Pfahringer}, \citenamefont {Witten},\ and\ \citenamefont
		{Trigg}}]{Weka:2009}%
	\BibitemOpen
	\bibfield  {author} {\bibinfo {author} {\bibfnamefont {E.}~\bibnamefont
			{Frank}}, \bibinfo {author} {\bibfnamefont {M.}~\bibnamefont {Hall}},
		\bibinfo {author} {\bibfnamefont {G.}~\bibnamefont {Holmes}}, \bibinfo
		{author} {\bibfnamefont {R.}~\bibnamefont {Kirkby}}, \bibinfo {author}
		{\bibfnamefont {B.}~\bibnamefont {Pfahringer}}, \bibinfo {author}
		{\bibfnamefont {I.~H.}\ \bibnamefont {Witten}}, \ and\ \bibinfo {author}
		{\bibfnamefont {L.}~\bibnamefont {Trigg}},\ }\bibfield  {title} {\enquote
		{\bibinfo {title} {Weka-a machine learning workbench for data mining},}\ }in\
	\href@noop {} {\emph {\bibinfo {booktitle} {Data mining and knowledge
				discovery handbook}}}\ (\bibinfo  {publisher} {Springer},\ \bibinfo {year}
	{2009})\ pp.\ \bibinfo {pages} {1269--1277}\BibitemShut {NoStop}%
	\bibitem [{\citenamefont {Hornik}, \citenamefont {Stinchcombe},\ and\
		\citenamefont {White}(1989)}]{Hornik:NN1989}%
	\BibitemOpen
	\bibfield  {author} {\bibinfo {author} {\bibfnamefont {K.}~\bibnamefont
			{Hornik}}, \bibinfo {author} {\bibfnamefont {M.}~\bibnamefont {Stinchcombe}},
		\ and\ \bibinfo {author} {\bibfnamefont {H.}~\bibnamefont {White}},\
	}\bibfield  {title} {\enquote {\bibinfo {title} {{Multilayer feedforward
					networks are universal approximators}},}\ }\href@noop {} {\bibfield
		{journal} {\bibinfo  {journal} {Neural Netw.}\ }\textbf {\bibinfo {volume}
			{2}},\ \bibinfo {pages} {359--366} (\bibinfo {year} {1989})}\BibitemShut
	{NoStop}%
	\bibitem [{\citenamefont {Gulli}\ and\ \citenamefont {Pal}(2017)}]{Gulli:book}%
	\BibitemOpen
	\bibfield  {author} {\bibinfo {author} {\bibfnamefont {A.}~\bibnamefont
			{Gulli}}\ and\ \bibinfo {author} {\bibfnamefont {S.}~\bibnamefont {Pal}},\
	}\href@noop {} {\emph {\bibinfo {title} {{Deep Learning with Keras}}}}\
	(\bibinfo  {publisher} {Packt Publishing},\ \bibinfo {address} {Birmingham},\
	\bibinfo {year} {2017})\BibitemShut {NoStop}%
	\bibitem [{\citenamefont {Bishop}(2006)}]{Bishop:book:2006}%
	\BibitemOpen
	\bibfield  {author} {\bibinfo {author} {\bibfnamefont {C.~M.}\ \bibnamefont
			{Bishop}},\ }\href@noop {} {\emph {\bibinfo {title} {Pattern Recognition and
				Machine Learning}}}\ (\bibinfo  {publisher} {Springer},\ \bibinfo {address}
	{Singapore},\ \bibinfo {year} {2006})\BibitemShut {NoStop}%
	\bibitem [{\citenamefont {Zhu}\ \emph {et~al.}(2018)\citenamefont {Zhu},
		\citenamefont {Mathai}, \citenamefont {Stevens}, \citenamefont {Verzicco},\
		and\ \citenamefont {Lohse}}]{Zhu:PRL2018}%
	\BibitemOpen
	\bibfield  {author} {\bibinfo {author} {\bibfnamefont {X.}~\bibnamefont
			{Zhu}}, \bibinfo {author} {\bibfnamefont {V.}~\bibnamefont {Mathai}},
		\bibinfo {author} {\bibfnamefont {R.~J. A.~M.}\ \bibnamefont {Stevens}},
		\bibinfo {author} {\bibfnamefont {R.}~\bibnamefont {Verzicco}}, \ and\
		\bibinfo {author} {\bibfnamefont {D.}~\bibnamefont {Lohse}},\ }\bibfield
	{title} {\enquote {\bibinfo {title} {{Transition to the Ultimate Regime in
					Two-Dimensional Rayleigh-B{\'e}nard Convection}},}\ }\href@noop {} {\bibfield
		{journal} {\bibinfo  {journal} {Phys. Rev. Lett.}\ }\textbf {\bibinfo
			{volume} {120}},\ \bibinfo {pages} {144502} (\bibinfo {year}
		{2018})}\BibitemShut {NoStop}%
	\bibitem [{\citenamefont {Iyer}\ \emph {et~al.}(2020)\citenamefont {Iyer},
		\citenamefont {Scheel}, \citenamefont {Schumacher},\ and\ \citenamefont
		{Sreenivasan}}]{Iyer:PNAS2020}%
	\BibitemOpen
	\bibfield  {author} {\bibinfo {author} {\bibfnamefont {K.~P.}\ \bibnamefont
			{Iyer}}, \bibinfo {author} {\bibfnamefont {J.~D.}\ \bibnamefont {Scheel}},
		\bibinfo {author} {\bibfnamefont {J.}~\bibnamefont {Schumacher}}, \ and\
		\bibinfo {author} {\bibfnamefont {K.~R.}\ \bibnamefont {Sreenivasan}},\
	}\bibfield  {title} {\enquote {\bibinfo {title} {{Classical 1/3 scaling of
					convection holds up to Ra $= 10^{15}$}},}\ }\href@noop {} {\bibfield
		{journal} {\bibinfo  {journal} {Proc. Natl. Acad. Sci. U.S.A.}\ }\textbf
		{\bibinfo {volume} {117}},\ \bibinfo {pages} {7594--7598} (\bibinfo {year}
		{2020})}\BibitemShut {NoStop}%
\end{thebibliography}
%

\end{document}